\documentclass[fleqn, usenatbib]{mnras}

\usepackage[T1]{fontenc}
\usepackage[utf8]{inputenc}

\usepackage{amsmath} 
\usepackage{amsfonts} 
\usepackage{amsbsy} 
\usepackage{amssymb} 
\usepackage{mathtools} 

\usepackage{graphicx} 
\usepackage{subcaption} 

\usepackage{acronym} 
\usepackage{soul} 
\usepackage{xcolor} 

\newacro{PPD}{proto-planetary disc}
\newcommand{\PPD}{\ac{PPD}}
\newacroplural{PPD}{proto-planetary discs}
\newcommand{\PPDs}{\acp{PPD}}

\newacro{RWI}{Rossby wave instability}
\newcommand{\RWI}{\ac{RWI}}

\newacro{SBI}{sub-critical baroclininc instability}
\newcommand{\SBI}{\ac{SBI}}

\newacro{COS}{convective over-stability}
\newcommand{\COS}{\ac{COS}}

\newacro{VSI}{vertical shear instability}
\newcommand{\VSI}{\ac{VSI}}

\newacro{ZVI}{zombie vortex instability}
\newcommand{\ZVI}{\ac{ZVI}}

\newacro{SI}{streaming instability}

\newacro{RDI}{resonant drag instability}

\newacroplural{RDI}{resonant drag instabilities}

\newacro{EI}{elliptical instability}
\newcommand{\EI}{\ac{EI}}

\newacro{ODE}{ordinary differential equation}

\newacroplural{ODE}{ordinary differential equations}

\newacro{PDE}{partial differential equation}

\newacroplural{PDE}{partial differential equations}

\newacro{TVA}{terminal velocity approximation}

\newacro{MMSN}{minimum mass solar nebulae}

\newcommand{\bb}[1]{\mbox{\boldmath{$#1$}}}

\newcommand{\bnabla}{\bb{\nabla}} 
\newcommand{\rd}{\mathrm{d}} 
\newcommand{\Dt}{\mathrm{D}_{t}} 

\newcommand{\bcdot}{\, \mbox{\boldmath{$\cdot$}} \,}

\newcommand{\bu}{\mathbf{u}} 

\newcommand{\e}{\mathrm{e}} 

\newcommand{\dtg}{\mu} 
\newcommand{\St}{\text{St}} 

\title[The effect of dust on vortices I]{The effect of dust on vortices I: Laminar models}
\author[N.~Magnan \& H.~Latter]{
    Nathan Magnan$^{1, 2}$,
    and Henrik N.~Latter$^{2}$\thanks{Contact e-mail: \href{mailto:hl278@cam.ac.uk}{hl278@cam.ac.uk}} 
    \\
    $^{1}$ Laboratoire Lagrange, Observatoire de la Côte d'Azur, Université de la Côte d'Azur, Nice, France\\
    $^{2}$ DAMTP, University of Cambridge, Cambridge, UK
    }

\begin{document}

\label{firstpage}
\pagerange{\pageref{firstpage}--\pageref{lastpage}}

\maketitle

\begin{abstract}
    One of the main questions regarding planet formation is how to cross the metre-scale barrier. Several theories rely on the formation of dust clumps dense enough to collapse under their own gravity. Vortices are promising candidate sites of clump formation because they can concentrate dust ‘laminarly’ by capturing particles, and ‘turbulently’ by creating the ideal conditions for the streaming instability. In this two-part series, we assess the validity of both pathways by investigating the effect of backreacting dust on vortices. This first paper focuses on the laminar pathway. We use multiple timescale analysis to create two models of vortex evolution. They differ in their assumptions regarding how much gas crosses the vortex’s boundary: the first one assumes that the vortex’s mass is constant, whereas the second one assumes that the gas density is constant. These two options epitomize the two ways in which vortices can respond to dust concentration. Essentially, as dust gets closer to the vortex centre, it loses angular momentum. To compensate, the gas must either move away from the vortex centre or change its vorticity (and therefore its shape). This choice neatly emerges from the conservation of a quantity akin to potential vorticity. Interestingly, we find that vortices that adjust their vorticity all evolve towards elliptically unstable shapes. And since the elliptical instability destroys the vortex, we conclude that dust imposes an upper bound on vortex lifetimes. If vortex destruction happens before the dust reaches the Hill density, the ‘laminar’ vortex pathway to planetesimal formation fails.
\end{abstract}

\begin{keywords}
    hydrodynamics --- protoplanetary discs --- planets and satellites: formation
    \vspace{-0.5 \baselineskip}
\end{keywords}

\section{Introduction}
\label{sec:Introduction}

Non-axisymmetric substructures have been observed by ALMA in several outer discs, and the VLTI may have detected one in the inner disc of HD 163296 \citep{Varga+2021, Gravity2021}. Since substructures can only be detected when the angular resolution of the instrument is sufficient, we can only do statistics on a small population of 37 well-resolved discs. 35 of those host substructures, and 8 of which are not axisymmetric \citep{Bae+2023}. This suggests that the proportion of \PPDs\ containing crescents may be as high as 20 \%.

Those crescents may be due to the lingering of dust and gas at the apocentre of the eccentric cavity carved by a binary \citep{Ragusa+2017}, the trapping of dust particles in the Lagrange
points of a companion \citep{Long+2022}, the intersection of a ring with a spiral wave \citep{Price+2018}, or the inner rim of an optically thick disc \citep{Ribas+2024}.

\phantom{\RWI, \SBI, \COS, \VSI, \ZVI}
\vspace{-2 \baselineskip}

But the current consensus is that most crescents are vortices. This interpretation is motivated by the fact that vortices are efficient dust traps \citep{BargeSommeria1995, AdamsWatkins1995, Tanga+1996, Chavanis2000}, and that many discs instabilities create large scale vortices: the Rossby wave instability (\RWI\ -- \citealt{Lovelace+1999, Li+2000}), the sub-critical baroclinic instability (\SBI\ -- \citealt{KlahrBodenheimer2003, Petersen+2007, LesurPapaloizou2010}), the convective over-stability (\COS\ -- \citealt{KlahrHubbard2014, Lyra2014, Latter2016}), the zombie vortex instability (\ZVI\ -- \citealt{Marcus+2013}), and perhaps the vertical shear instability (\VSI\ -- \citealt{UrpinBrandenburg1998, Nelson+2013, Richard+2016, Lesur+2025}).

Because they capture and concentrate pebbles, vortices are sometimes seen as `planetesimal factories'. The idea is that the dust density at the vortex's centre increases over time, so it must eventually become larger than the Hill density. The core must then collapse gravitationally into planetesimals.

Unfortunately, we do not know what is the maximal dust density that vortices can induce. \cite{LyraLin2013} argue that if there are small-scale turbulent eddies within a vortex, they will diffuse the dust radially. This enables a steady state where the outward flux due to diffusion compensates the inward flux due to dust~capture. 

Alternatively, dust concentration may stop due to a laminar mechanism. Indeed, as dust concentrates, the dust-to-gas ratio in the vortex's core increases. When it reaches unity, the dust's feedback onto the gas becomes relevant, and may affect the flow in such a way as to inhibit the vortex's ability to further concentrate dust. 

We are not the first to study this possiblity. For instance, \cite{Surville+2016} built an analytical toy model predicting that dust density does indeed saturate. Unfortunately, it only applies to a single point at the centre of the vortex. 

The question has also been tackled with numerical experiments, but those disagree on several important points:
\begin{itemize}
    \vspace{-0.5 \baselineskip}
    \item Starting from similar metallicities and working with similar-sized particles, \cite{Fu+2014} and \cite{SurvilleMayer2019} find that it takes a hundred orbits to reach a dust-to-gas ratio of order unity, whereas \cite{Meheut+2012b} and \cite{Raettig+2015} report that a dozen orbits are sufficient.
    \item \cite{CrnkovicRubsamen+2015}, \cite{Surville+2016} and \cite{Miranda+2017} all find that as the vortex captures more and more dust, its core loses vorticity. However, there is some tension regarding this vortex damping process. Firstly, \cite{CrnkovicRubsamen+2015} reports that it only occurs for large particles, whereas the other teams report it for all particles. Furthermore, \cite{Surville+2016} predict that the vortex’s core is completely damped and stops concentrating dust when the dust-to-gas ratio reaches 0.5, whereas \cite{Miranda+2017} reach a dust-to-gas ratio well above unity.
\end{itemize}

The goal of the present paper is to make progress on the analytical front. We use the shearing box and multiple timescale analysis to study the effect of small-Stokes-number particles on small anticyclonic Kida vortices. 

The structure of the paper is as follows. We start by presenting in \S\ref{sec:Governing_equations} the physical system and its governing equations, and in  \S\ref{sec:Kida_vortex} the gaseous vortex in which dust will accumulate. Then in \S\ref{sec:Model_0} we use the regime of test particles as a simple setting in which to introduce our mathematical methods. At this stage, we are ready to consider the regime of massive particles and to show how dust affects vortex evolution (\S\ref{sec:Model_AB}). Finally, we discuss our hypotheses and our results in \S\ref{sec:Discussion}, before concluding in \S\ref{sec:Conclusion}.

\vspace{-1 \baselineskip}
\section{Governing equations}
\label{sec:Governing_equations}

\newcommand{\rhog}{\rho_{g}} 
\newcommand{\bug}{\bu_{g}} 
\newcommand{\rhod}{\rho_{d}} 
\newcommand{\bud}{\bu_{d}} 

We consider a gas and dust mixture, which we model as a two-fluid system. The gas is described by its density~$\rhog$, its velocity~$\bug$ and its pressure~$P$. Our model screens out sound waves, but allows $\rhog$ to depend slowly on time -- more details in \S\ref{sec:gas_model}. The dust is pressure-less, so it is described by its density~$\rhod$ and velocity~$\bud$. Crucially, we assume that $\rhog$ and $\rhod$ are uniform. This will be justified later, in \S\ref{sub:Model_0_model}.

The two fluids are coupled by drag and its back-reaction. We assume that the relative velocities between dust and gas are small enough that we are either in the Epstein or Stokes regime, depending on the size of the dust particles. We also assume that all the particles have the same size. This allows us to introduce a universal stopping time $\tau$.

\newcommand{\eX}{\ensuremath{\mathbf{e}_{X}}} 
\newcommand{\eY}{\ensuremath{\mathbf{e}_{Y}}} 
\newcommand{\eZ}{\ensuremath{\mathbf{e}_{Z}}} 

To account for the \PPD\ context, we work in the shearing box \citep{GoldreichLyndenBell1965, Hawley1995, LatterPapaloizou2017}. The centre of the box orbits at frequency $\Omega$ and radius $r_{0}$. The local Cartesian coordinate system has its $X$-axis oriented in the radial direction, its $Y$-axis in the azimuthal direction, and its $Z$-axis in the direction normal to the disc's plane. The Keplerian flow, the tidal force, and the Coriolis force take their standard form,
\begin{subequations}
    \label{eq:flow_and_forces_in_shearing_box}
    \begin{align}
        \bu_{\mathrm{sb}} &= - S X \eY , \label{eq:flow_in_shearing_box} \\
        - \bnabla \Phi_{t} &= 2 \Omega S X \eX , \label{eq:forces_in_shearing_box_tidal_potential} \\
        \mathbf{f}_{\text{Co}} (\bu) &= - 2 \Omega \, \eZ \wedge \bu , \label{eq:forces_in_shearing_box_Coriolis}
    \end{align}
\end{subequations}
where the subscript $\mathrm{sb}$ stands for \textit{shearing box}, ${ S \approx (3/2) \, \Omega }$ is the local shearing rate of the disc, and $\bu$ could represent either the gas or the dust's velocity. Note that we ignore the disc's large-scale pressure gradient and the vertical component of gravity. Indeed, our focus is on the midplane layer.

All of this is summed up in the equations
\begin{subequations}
    \label{eq:governing_equations_standard_coordinates}
    \begin{align}
        & \partial_{t} \ln{(\rhog)} = - \bnabla \bcdot \bug , \phantom{\frac{1}{1}} \label{eq:governing_equations_standard_coordinates_gas_density} \\ 
        & \partial_{t} \ln{(\rhod)} = - \bnabla \bcdot \bud , \phantom{\frac{1}{1}} \label{eq:governing_equations_standard_coordinates_dust_density} \\
        & \Dt \bug = - \frac{\bnabla P}{\rhog} - 2 \Omega \, \eZ \wedge \bug + 2 \Omega S X \eX + \frac{\rhod}{\rhog} \frac{\bud - \bug}{\tau} , \label{eq:governing_equations_standard_coordinates_gas_momentum} \\
        & \Dt \bud = - 2 \Omega \, \eZ \wedge \bud + 2 \Omega S X \eX - \frac{\bud - \bug}{\tau} , \label{eq:governing_equations_standard_coordinates_dust_momentum}
    \end{align}
\end{subequations}
where ${ \Dt }$ is the advective derivative. Note that just like in the incompressible model, $P$ is a Lagrange multiplier. The only difference is that instead of the enforcing the constraint ${ \bnabla \bcdot \bug = 0 }$, it enforces ${ \bnabla \bcdot \bug = - \partial_{t} \ln{(\rhog)} }$. Indeed, ${ \partial_{t} \ln{(\rhog)} }$ is not a variable but a parameter of our model.

\newcommand{\bv}{\mathbf{v}} 

\newcommand{\fg}{f_{g}} 
\renewcommand{\fd}{f_{d}} 

\newcommand{\bGi}{\mathbf{G}_{1}} 
\newcommand{\bGii}{\mathbf{G}_{2}} 

We shall sometimes use the variables ${ \bu = \fg \, \bug + \fd \, \bud }$, ${ \bv = \bud - \bug }$, ${ \rho = \rhog + \rhod }$ and ${ \dtg = \rhod / \rhog }$. In the Eulerian view, $\bu$ is the velocity of the centre-of-mass of two colocated gas and dust fluid parcels, $\bv$ is the relative velocity between those two parcels, $\rho$ is the total density at the colocation point, and $\dtg$ is the dust-to-gas density ratio. ${ \fg = \rhog / \rho }$ and ${ \fd = \rhod / \rho }$ are the gas and dust mass fractions. Since our model allows the gas density to vary in time, our governing equations could be more complex than those of \cite{YoudinGoodman2005}. Thankfully, since $\rhog$ and $\rhod$ are both uniform, we simply get
\begin{subequations}
    \label{eq:governing_equations}
    \begin{align}
        & \partial_{t} \ln{(\rho)} = - \bnabla \bcdot \bu , \phantom{\frac{1}{1}} \label{eq:governing_equations_total_density} \\ 
        & \partial_{t} \ln{(\dtg)} = - \bnabla \bcdot \bv , \phantom{\frac{1}{1}} \label{eq:governing_equations_dust_to_gas_ratio} \\
        & \partial_{t} \bu + \bu \bcdot \bnabla \bu + \bGi = - \fg \bnabla h - 2 \Omega \, \eZ \wedge \bu + 2 \Omega S X \eX , \phantom{\frac{1}{1}} \!\!\!\!\!\! \label{eq:governing_equations_centre_of_mass} \\
        & \partial_{t} \bv + \bu \bcdot \bnabla \bv + \bv \bcdot \bnabla \bu + \bGii = \bnabla h \! - \! 2 \Omega \, \eZ \wedge \bv \! - \! \frac{1 \! + \! \dtg}{\tau} \bv , \!\!\! \label{eq:governing_equations_relative_velocity}
    \end{align}
\end{subequations}
where ${h = P / \rhog}$ is a variable equivalent to pressure and
\begin{subequations}
    \label{eq:advective_mess}
    \begin{align}
        \bGi &= \fg \fd \left[ (\bnabla \bcdot \bv) \, \bv + \bv \bcdot \bnabla \bv \right] , \label{eq:advective_mess_1} \\
        \bGii &= \left(\fg^{2} - \fd^{2}\right) \bv \bcdot \bnabla \bv . \label{eq:advective_mess_2}
    \end{align}
\end{subequations}
These `barycentric' variables are more convenient than the standard ones because only one equation, \eqref{eq:governing_equations_relative_velocity}, needs to be regularised in the limit of short stopping times.

\section{The Kida vortex}
\label{sec:Kida_vortex}

\newcommand{\buK}{\bu_{\text{K}}} 
\newcommand{\hK}{h_{\text{K}}} 

Our strategy is to start from a gas-only flow, to add dust, and to see how that affects the flow. So the first thing we need is a model for gaseous vortices in \PPDs. We shall follow \cite{Chavanis2000} and \cite{LesurPapaloizou2009} and use the Kida model. This is an elliptical patch of aspect ratio $\alpha$ and constant vorticity~${ \omega_{\text{v}} }$ embedded in a shear flow of rate~$S$.\footnote{Note that the total vorticity is then ${ \omega_{\text{t}} = \omega_{\text{v}} - S }$.} For this to be an exact solution to the inviscid Navier-Stokes equations, the ellipse generally needs to stretch and rotate over time \citep{Kida1981}. But steady state is possible if
\begin{equation}
    \label{eq:steady_Kida_vortices}
    \frac{\omega_{\text{v}}}{S} = - \frac{1}{\alpha} \left( \frac{\alpha + 1}{\alpha - 1} \right) .
\end{equation} 
We shall focus on the anticyclonic steady states, because they can trap dust. The flow inside the ellipse is then simply
\begin{subequations}
    \label{eq:Kida_vortex}
    \begin{align}
        \buK &= \frac{S}{\alpha - 1} \left( \alpha^{-1} Y \, \eX - \alpha X \, \eY \right) , \label{eq:Kida_vortex_u} \\
        \hK &= \frac{S}{\alpha - 1} \bigg[ \left( \frac{S / 2}{\alpha - 1} - \Omega \right) X^{2} + \left( \frac{S / 2}{\alpha - 1} - \frac{\Omega}{\alpha} \right) Y^{2} \bigg] , \label{eq:Kida_vortex_P}
    \end{align}
\end{subequations}
where the subscript $\text{K}$ stands for \textit{Kida}, not \textit{Keplerian}.

\vspace{-0.5 \baselineskip}
\section{Test particles}
\label{sec:Model_0}

In order to build intuition, let us start in the regime of test particles, ${ \dtg = 0 }$. This removes the backreaction, so the gas follows the Kida flow and we can study how dust responds to vortices without having to worry about vortex evolution. It also allows us to present in a simple setting the mathematical tools that we
will use to derive full models in~\S\ref{sec:Model_AB}.

\renewcommand{\tt}{\tilde{t}} 

\newcommand{\bX}{\mathbf{X}} 
\newcommand{\tbX}{\tilde{\bX}} 
\newcommand{\tX}{\tilde{X}} 
\newcommand{\tY}{\tilde{Y}} 
\newcommand{\tbnabla}{\tilde{\bnabla}} 

\newcommand{\trhod}{\tilde{\rho}_{d}} 

\newcommand{\tbud}{\tilde{\bu}_{d}} 
\newcommand{\tbuK}{\tilde{\bu}_{\text{K}}} 

We adimensionalise the dust equations using an arbitrary lengthscale $L$, the orbital timescale ${ T = 1 / \Omega }$ and an arbitrary dust density scale ${ \overline{\rho}_{d} }$. The time coordinate is replaced by ${ \tt = \Omega t }$, the spatial coordinate is replaced by ${ \tbX = \bX / L }$, the dust velocity is replaced by ${ \tbud = \bud / (L \Omega) }$, the dust density is replaced by ${ \trhod = \rhod / \overline{\rho}_{d} }$, and Eqs.~(\ref{eq:governing_equations_standard_coordinates_dust_density}, \ref{eq:governing_equations_standard_coordinates_dust_momentum}, \ref{eq:Kida_vortex_u}) become
\begin{small}
    \begin{subequations}
        \label{eq:test_particles_adimensional_equations}
        \begin{align}
            & \partial_{\tt} \ln{(\trhod)} = - \tbnabla \bcdot \tbud , \phantom{\frac{1}{1}} \label{eq:test_particles_adimensional_equations_continuity} \\
            & \partial_{\tt} \, \tbud + \tbud \bcdot \tbnabla \tbud = - 2 \, \eZ \wedge \bud + 2 (S / \Omega) \tX \eX \! - \! \frac{1}{\St} (\tbud \! - \! \tbuK) , \!\!\! \label{eq:test_particles_adimensional_equations_momentum} \\
            & \tbuK = \frac{S / \Omega}{\alpha - 1} \left( \alpha^{-1} \tY \,  \eX - \alpha \tX \, \eY \right) , \label{eq:test_particles_adimensional_equations_Kida_velocity}
        \end{align}
    \end{subequations}
\end{small}
\!\!This shows that the system is controlled by three parameters: the disc's shear rate ${ S / \Omega }$, the vortex's aspect ratio $\alpha$, and the particles' Stokes number ${ \St = \Omega \tau }$.

\vspace{-0.5 \baselineskip}
\subsection{Multiple timescale analysis}
\label{sub:Multiple_timescale_analysis}

If the particles are well coupled, we expect the system to exhibit dynamics on three well-separated timescales: the dust velocity should relax towards that of the gas quickly, parcels should go around the vortex on the reference (\textit{i.e.} orbital) timescale, and the dust density should evolve slowly.

\newcommand{\trhodo}{\tilde{\rho}_{d, 0}}
\newcommand{\trhodi}{\tilde{\rho}_{d, 1}}
\newcommand{\trhodii}{\tilde{\rho}_{d, 2}}

\newcommand{\tbudo}{\tilde{\bu}_{d, 0}}
\newcommand{\tbudi}{\tilde{\bu}_{d, 1}}
\newcommand{\tbudii}{\tilde{\bu}_{d, 2}}

To leverage this insight, we use a multiple timescale analysis (\citealt{BenderOrszag1999}, chapter 11, section 2). We first expand each variable in powers of $\St$,
\begin{align}
    \trhod &= \trhodo + \St \, \trhodi + \St^{2} \, \trhodii + \text{ ..,} \nonumber\\ 
    \tbud &= \tbudo + \St \, \tbudi + \St^{2} \, \tbudii + \text{ ...} \nonumber
\end{align}
Each of the variables ${\tilde{f}_{i}}$ should be of order 1 and remains so at all times, otherwise the asymptotic ordering breaks down. 

The second step is to model each variable as being a function of several time variables $\tt_{1}$, $\tt_{2}$, $\tt_{3}$, ... rather than a single, universal time~$\tt$. We can then replace \smash{${ \partial_{\tt} }$} by \smash{${ \partial_{\tt_{1}} \!\! + \partial_{\tt_{2}} \!\! + \partial_{\tt_{3}} \!\! + ... }$} and assume that \smash{${ \partial_{\tt_{n + 1}} \tilde{f}_{i} }$} appears one order later than \smash{${ \partial_{\tt_{n}} \tilde{f}_{i} }$} in the expansion in powers of $\St$. Essentially, $\tt_{1}$ represents what happens on short timescales, $\tt_{2}$ what happens on intermediate timescales, $\tt_{3}$ what happens on long timescales,~\textit{etc}.

Finally, it seems reasonable to interpret $\tt_{1}$ as the drag timescale, $\tt_{2}$ as the orbital timescale and $\tt_{3}$ the dust concentration timescale. We shall therefore assume that \smash{$\partial_{\tt_{2}} \tilde{f}_{i}$} is of the same order as \smash{$\tilde{f}_{i}$}. The overall substitution is then
\begin{equation}
    \nonumber
    \partial_{\tt} \tilde{f}_{i} \rightarrow \St^{-1} \, \partial_{\tt_{1}} \tilde{f}_{i} + \partial_{\tt_{2}} \tilde{f}_{i} + \St \, \partial_{\tt_{3}} \tilde{f}_{i} .
\end{equation}
There may be other consistent scalings, but they would beyond the scope of this paper.

\subsection{Leading order}
\label{sub:Model_0_leading_order}

At order $\St^{-1}$, the momentum equation~\eqref{eq:test_particles_adimensional_equations_momentum} simplifies to
\begin{equation}
    \label{eq:Model_0_leading_order_momentum}
    \partial_{\tt_{1}} \tbudo + \tbudo = \tbuK .
\end{equation}
This indicates that whatever the initial state, the leading-order dust velocity relaxes to the Kida velocity on the `fast' timescale. Therefore, to an observer living on one of the longer timescales, $\tbudo$ appears equal to $\tbuK$ at all times.

The initial relaxation period is of little interest to us, so to simplify our model we select the initial condition ${ \bud (t = 0) = \buK }$. Thanks to this constraint, the equality ${ \tbudo = \tbuK }$ hold on the fast timescale as well.

The leading-order continuity equation~\eqref{eq:test_particles_adimensional_equations_continuity} is ${ \partial_{\tt_{1}} \trhodo = 0 }$.

\subsection{Second order}
\label{sub:Model_0_second_order}

At order $\St^{0}$, the continuity equation~\eqref{eq:test_particles_adimensional_equations_continuity} becomes
\begin{equation}
    \label{eq:Model_0_second_order_continuity}
    \partial_{\tt_{2}} \, \trhodo + \partial_{\tt_{1}} \, \trhodi = - \trhodo \, \bnabla \bcdot \tbudo .
\end{equation}
The first and last terms are independent of $\tt_{1}$, so ${ \partial_{\tt_{1}} \trhodi }$ must be independent of $\tt_{1}$ as well. To an observer living on the `fast' timescale, this derivative appears constant. If this constant was not zero, then $\trhodi$ would appear to grow linearly in time, so ${ \St \, \trhodi}$ would eventually become comparable to ${ \trhodo }$, and the asymptotic ordering would break down. Therefore, we impose ${ \partial_{\tt_{1}} \trhodi = 0 }$. Formally, this is the `non-secularity' condition imposed by \cite{BenderOrszag1999} via their Eq.~(11.2.8). It just takes a simpler form here.

Now since $\tbudo$ is equal to $\tbuK$, and since $\tbuK$ is divergence-free, $\trhodo$ must be independent of~$\tt_{2}$. This confirms our intuition from \S\ref{sub:Multiple_timescale_analysis} that the dust density only evolves on the `slow' timescale $\tt_{3}$.

\renewcommand{\th}{\tilde{h}} 
\newcommand{\thK}{\tilde{h}_{K}} 

\newcommand{\tDelta}{\tilde{\Delta}} 

Since $\tbudo$ is independent of $\tt_{2}$, the second-order dust momentum equation writes
\begin{equation}
    \partial_{\tt_{1}} \tbudi + \tbudi = - \tbuK \bcdot \tbnabla \tbuK - 2 \, \eZ \wedge \tbuK + 2 (S/\Omega) \tX \eX . \nonumber
\end{equation}
At this stage, one should remember that $\buK$ and $\hK$ form an exact and steady solution to the Navier-Stokes equations. This allows us to simplify the right hand side to
\begin{equation}
    \label{eq:Model_0_second_order_momentum}
    \partial_{\tt_{1}} \tbudi + \tbudi = \tbnabla \thK ,
\end{equation}
where ${ \th = h / (L \Omega)^{2} }$ is the adimensional pseudo-enthalpy.

This equation indicates that, whatever the initial state, $\tbudi$ relaxes to $\tbnabla \thK$ after a few dust stopping times. As expected, small particles mostly follow the gas, except for a slow drift towards pressure maxima.

\subsection{Third order}
\label{sub:Model_0_third_order}

At order $\St^{1}$, the continuity equation becomes
\begin{equation}
    \partial_{\tt_{3}} \, \trhodo + \partial_{\tt_{2}} \, \trhodi + \partial_{\tt_{1}} \, \trhodii = - \trhodo \, \tDelta \thK , \nonumber
\end{equation}
where $\tDelta$ is the adimensional Laplacian. As before, our asymptotic ordering is only consistent with ${ \partial_{\tt_{2}} \, \trhodi = \partial_{\tt_{1}} \, \trhodii = 0 }$. This leads to
\begin{equation}
    \label{eq:Model_0_third_order_continuity}
    \partial_{\tt_{3}} \ln{(\trhodo)} = - \tDelta \thK .
\end{equation}

\subsection{Model 0}
\label{sub:Model_0_model}

In dimensional terms, Eqs.~\eqref{eq:Model_0_leading_order_momentum}, \eqref{eq:Model_0_second_order_momentum} and~\eqref{eq:Model_0_third_order_continuity} translate to
\begin{subequations}
    \label{eq:Model_0}
    \begin{align}
        \bud (t) &\approx \buK + \tau \bnabla \hK , \label{eq:Model_0_dust_velocity} \\
        \rhod (t) &\approx \rhod (t = 0) \, \e^{- \tau \, \Delta \hK \, t} . \label{eq:Model_0_dust_density}
    \end{align}
\end{subequations}
This is our first approximate vortex model, valid in the regime of well-coupled test particles. We shall call it `model 0'.

Note that the $e$-folding rate of the dust density,
\begin{equation}
    \label{eq:Model_0_dust_capture_rate}
    \tau \, |\Delta \hK| = 2 \, \St \, (S / \Omega) \, \frac{\alpha^{2} - (S / \Omega) \, \alpha - 1}{\alpha \, (\alpha - 1)^{2}} \, \Omega ,
\end{equation} 
is in perfect agreement with Eq.~(24) from \cite{Chavanis2000}. Indeed, they compute the trajectory of Lagrangian particles and find that their distance to the vortex's centre decreases exponentially over time, with an $e$-folding rate that is half of ours. That makes sense: if the trajectory of single particles spiraling into the vortex scales with ${ \e^{- t / t_{\text{capt}}} }$, then the area of the ellipse of aspect ratio $\alpha$, centered on the vortex's center, and whose boundary tracks a particle, will scale with ${ \e^{- 2 t / t_{\text{capt}}} }$. Therefore, the $e$-folding rate of Eulerian dust density is twice the $e$-folding rate of Lagrangian trajectories.

Note also that this $e$-folding rate is uniform. This is due to a peculiarity of Kida's vortex: its pressure Laplacian is uniform. But it is important, as this is what allows us to assume that the dust density is uniform. Indeed, if we had included the advection term in the dust continuity equations, we would have found
\begin{equation}
    \nonumber
    \partial_{t} \rhod + \bud \bcdot \bnabla \rhod = - \rhod \, \St \, \Delta \hK .
\end{equation}
This shows that if the initial dust density is uniform, it remains so at all times. In other words, the linearity of Kida's vortex ensures the existence of uniform solutions, and the goal of the present paper is to approximate those.

\section{Massive particles}
\label{sec:Model_AB}

\newcommand{\trho}{\tilde{\rho}}
\newcommand{\trhoo}{\tilde{\rho}_{0}}
\newcommand{\trhoi}{\tilde{\rho}_{1}}
\newcommand{\trhoii}{\tilde{\rho}_{2}}

\newcommand{\dtgo}{\dtg_{0}}
\newcommand{\dtgi}{\dtg_{1}}
\newcommand{\dtgii}{\dtg_{2}}

\newcommand{\tho}{\tilde{h}_{0}}
\newcommand{\thi}{\tilde{h}_{1}}
\newcommand{\thii}{\tilde{h}_{2}}

\newcommand{\tbu}{\tilde{\bu}}
\newcommand{\tbuo}{\tilde{\bu}_{0}}
\newcommand{\tbui}{\tilde{\bu}_{1}}
\newcommand{\tbuii}{\tilde{\bu}_{2}}

\newcommand{\tbv}{\tilde{\bv}}
\newcommand{\tbvo}{\tilde{\bv}_{0}}
\newcommand{\tbvi}{\tilde{\bv}_{1}}
\newcommand{\tbvii}{\tilde{\bv}_{2}}

Let us now leave the regime of test particles, and see how the dust affects vortex evolution. To do so, we shall use the same multiple-timescale method as in \S\ref{sec:Model_0}, but applied to the barycentric variables $\rho$, $\dtg$, $h$, $\bu$ and $\bv$. We use the same set of timescales, with the same interpretation. To make sure that the magnitude of every term is entirely determined by $\St$, we must assume that the dust-to-gas ratio is not much larger than one. We implicitly made the same assumption regarding $\alpha$ and $S / \Omega$ in the previous section.


\subsection{Leading order}
\label{sub:Model_AB_leading_order}

\subsubsection{Relative velocity}
\label{ssub:Model_AB_leading_order_relative_velocity}

At order $\St^{-1}$, the relative velocity equation~\eqref{eq:governing_equations_relative_velocity} writes
\begin{equation}
    \label{eq:Model_AB_leading_order_relative_velocity}
    \partial_{\tt_{1}} \, \tbvo + (1 + \dtgo) \, \tbvo = \mathbf{0} .
\end{equation}
Just like in the test particle regime, $\tbvo$ converges to zero on the fast timescale, whatever the initial conditions. Therefore, to an observer living on the orbital or slow timescale, $\tbvo$ appears null at all times. 

As before, we select the initial condition ${ \bv (0) = \mathbf{0} }$ so that ${ \tbvo = \mathbf{0} }$ at all times and on all timescales. This filters out the initial relaxation phase, but that is acceptable since our interest is in the long-term dynamics.

\subsubsection{Other variables}
\label{ssub:Model_AB_leading_order_centre_of_mass_velocity}

The leading-order equations for the other variables are
\begin{subequations}
    \label{eq:Model_AB_leading_order}
    \begin{align}
        \partial_{\tt_{1}} \tbuo &= \mathbf{0} , \label{eq:Model_AB_leading_order_centre_of_mass_velocity} \\
        \partial_{\tt_{1}} \ln{(\trhoo)} &= 0 , \label{eq:Model_AB_leading_order_total_density} \\
        \partial_{\tt_{1}} \ln{(\dtgo)} &= 0 . \label{eq:Model_AB_leading_order_dust_to_gas_ratio}
    \end{align}
\end{subequations}
In short, those variables are not necessarily null like $\tbvo$, but they are independent of $\tt_{1}$.

\subsection{Second order}
\label{sub:Model_AB_second_order}

\subsubsection{Centre-of-mass velocity}
\label{ssub:Model_AB_second_order_centre_of_mass_velocity}

At order $\St^{0}$, the barycentre equation~\eqref{eq:governing_equations_centre_of_mass} writes\footnote{Formally, there is an extra term in ${ \partial_{\tt_{1}} \tbudi }$, but the asymptotic ordering breaks down if this term is non-zero, so we assume that $\tbudi$ is independent of $\tt_{1}$.}
\begin{equation}
    \label{eq:Model_AB_second_order_centre_of_mass_velocity}
    \partial_{\tt_{2}} \, \tbuo + \tbuo \bcdot \tbnabla \tbuo = - \fg \tbnabla \tho - 2 \Omega \, \eZ \wedge \tbuo + 2 (S / \Omega) \tX \eX .
\end{equation}
We recognise the Navier-Stokes equation in the shearing box. As such, we know that many exact solutions exists. But since we want to study how dust affects Kida vortices, we impose ${ \tbuo = \tbuK }$ and ${ \tho = (1 + \dtgo) \, \thK }$ for some aspect ratio $\alpha$, which may or may not vary on the slow timescale $\tt_{3}$. In other words: we force the centre-of-mass to follow the Kida flow, but we leave open the possibility of this Kida flow slowly evolving.

\subsubsection{Relative velocity}
\label{ssub:Model_AB_second_order_relative_velocity}

At second-order, the relative-velocity equation becomes
\begin{equation}
    \label{eq:Model_AB_second_order_relative_velocity}
    \partial_{\tt_{1}} \, \tbvi + (1 + \dtgo) \, \tbvi = (1 + \dtgo) \, \tbnabla \thK . 
\end{equation}
This shows that $\tbvi$ quickly converges to ${ \tbnabla \thK }$. Physically, this means that the dust slowly drifts towards Kida pressure maxima, just like in \S\ref{sec:Model_0}.

\subsubsection{Other variables}
\label{ssub:Model_AB_second_order_other}

At order $\St^{0}$, the density equations~(\ref{eq:governing_equations_total_density}, \ref{eq:governing_equations_dust_to_gas_ratio}) become
\begin{subequations}
    \label{eq:Model_AB_second_order}
    \begin{align}
    \partial_{\tt_{2}} \trhoo + \partial_{\tt_{1}} \trhoi &= - \trhoo \, \tbnabla \bcdot \tbuo = 0 , \label{eq:Model_AB_second_order_total_density} \\
    \partial_{\tt_{2}} \dtgo + \partial_{\tt_{1}} \dtgi &= - \dtgo \, \tbnabla \bcdot \tbvo = 0 . \label{eq:Model_AB_second_order_dust_to_gas_ratio}
    \end{align} 
\end{subequations}
Just like with Eq.~\eqref{eq:Model_0_second_order_continuity}, we can only maintain the asymptotic ordering if $\trhoi$ and $\dtgi$ are independent of $\tt_{1}$, and if $\trhoo$ and $\dtgo$ are independent of $\tt_{2}$.

\subsection{Third order}
\label{sub:Model_A_third_order}

\subsubsection{Dust-to-gas ratio}
\label{ssub:Model_AB_third_order_dust_to_gas_ratio}

At order $\St^{1}$, the dust-to-gas ratio equation~\eqref{eq:governing_equations_dust_to_gas_ratio} writes
\begin{equation}
    \nonumber
    \partial_{\tt_{3}} \dtgo + \partial_{\tt_{2}} \dtgi + \partial_{\tt_{3}} \dtgii = - \dtgo \, \tbnabla \bcdot \tbvi . 
\end{equation}
Just like with Eq.~\eqref{eq:Model_0_third_order_continuity}, we must impose ${ \partial_{\tt_{2}} \dtgi = \partial_{\tt_{1}} \dtgii = 0 }$. We get
\begin{equation}
    \label{eq:Model_AB_third_order_dust_to_gas_ratio}
    \partial_{\tt_{3}} \ln{(\dtgo)} = - \tDelta \thK . 
\end{equation}
At this stage, the point we made at the end of \S\ref{ssub:Model_AB_second_order_centre_of_mass_velocity} becomes relevant: $\thK$ is the Kida pressure for a certain aspect ratio $\alpha$, and $\alpha$ may depend on $\tt_{3}$, so Eq.~\eqref{eq:Model_AB_third_order_dust_to_gas_ratio} does not form a closed problem: we need one more equation for $\alpha$.

\subsubsection{A conserved quantity}
\label{ssub:Model_AB_third_order_conserved_quantity}

The most elegant way to predict the evolution of $\alpha$ is to leverage a conserved quantity of the dust-gas system. This hidden constant is related to potential vorticity, so the simplest way to exhibit it starts from the vorticity equation.

\newcommand{\tomegao}{\tilde{\omega}_{0}}
\newcommand{\tomegai}{\tilde{\omega}_{1}}
\newcommand{\tomegaii}{\tilde{\omega}_{2}}

\newcommand{\TV}{\Theta}

Taking the curl of the centre-of-mass-velocity equation~\eqref{eq:governing_equations_centre_of_mass} and projecting along $\eZ$ gives
\begin{equation}
    \label{eq:Helmoltz_equation}
    \partial_{t} \omega + \bu \bcdot \bnabla \omega + \left( \bnabla \wedge \bGi \right)_{Z} = - (\omega + 2 \Omega) (\bnabla \bcdot \bu) ,
\end{equation}
where ${ \omega = \left( \bnabla \wedge \bu \right)_{Z} }$ is the vertical centre-of-mass vorticity. Now since Kida's flow is incompressible and has uniform vorticity, this equation simplifies at order $\St^{1}$ to
\begin{small}
    \begin{equation}
        \nonumber
        \partial_{\tt_{3}} \tomegao + \partial_{\tt_{2}} \tomegai + \partial_{\tt_{1}} \tomegaii + \tbuo \bcdot \tbnabla \tomegai = - (\tomegao + 2) (\tbnabla \bcdot \tbui) .
    \end{equation}
\end{small}
\!\!\!Since $\tomegao$ is constant on the fast timescale, we cannot let $\tomegaii$ grow on that timescale and must impose ${ \partial_{\tt_{1}} \tomegaii = 0 }$. This leaves
\begin{equation}
    \nonumber
    \partial_{\tt_{2}} \tomegai + \tbuo \bcdot \tbnabla \tomegai = - (\tomegao + 2) (\tbnabla \bcdot \tbui) - \partial_{\tt_{3}} \tomegao .
\end{equation}
The right hand side is uniform, so if $\tomegai$ is initially uniform, it remains so at all times. This allows us to drop the advective term on the left hand side to get
\begin{equation}
    \nonumber
    \partial_{\tt_{3}} \tomegao + \partial_{\tt_{2}} \tomegai = - (\tomegao + 2) (\tbnabla \bcdot \tbui) .
\end{equation}
Using the usual argument, we conclude that $\tomegai$ does not evolve on the orbital timescale. This leads to
\begin{equation}
    \label{eq:Model_AB_third_order_specific_angular_momentum}
    \partial_{\tt_{3}} \TV + \TV \, \tbnabla \bcdot \tbui = 0 ,
\end{equation}
where ${ \TV = \tomegao + 2 }$ is a total vorticity. It combines two contributions: a local one due to the vortex~($\omega$), and a global one due to the disc~($2 \Omega$).

The next step is to eliminate ${ \tbnabla \bcdot \tbui }$. This is pretty straightforward. Indeed, at third order, Eq.~\eqref{eq:governing_equations_total_density} becomes
\begin{equation}
    \nonumber
    \partial_{\tt_{3}} \trhoo + \partial_{\tt_{2}} \trhoi + \partial_{\tt_{1}} \trhoii = - \trhoo \, \tbnabla \bcdot \tbui .
\end{equation}
As usual, ${ \partial_{\tt_{2}} \trhoi }$ and ${ \partial_{\tt_{1}} \trhoii }$ must be null, so
\begin{equation}
    \label{eq:Model_AB_third_order_total_density}
    \tbnabla \bcdot \tbui = - \partial_{\tt_{3}} \ln{(\trhoo)} ,
\end{equation}
Note that as stated in \S\ref{sec:Governing_equations}, ${ \partial_{\tt_{3}} \ln{(\trhoo)} }$ is not a new variable but a parameter of our model.

\newcommand{\PV}{\theta}

Finally, we can combine Eqs.~\eqref{eq:Model_AB_third_order_specific_angular_momentum} and~\eqref{eq:Model_AB_third_order_total_density} to get
\begin{equation}
    \label{eq:conservation_of_potential_vorticity}
    \partial_{\tt_{3}} \PV = 0 ,
\end{equation}
where ${ \PV = \TV / \trhoo }$ is a sort of potential vorticity for the ${\{\text{dust} + \text{gas} \}}$ system. We suspect that the reason it is conserved has to do with the analogy between thermodynamics and dust dynamics found by \cite{LinYoudin2017}.

From there, it is easy to determine the evolution of $\alpha$. Indeed, we prescribe ${ \trhoo (\tt_{3}) }$ and ${ \tomegao (0) }$. The conservation of potential vorticity gives us ${ \tomegao (\tt_{3}) }$, and Eq.~\eqref{eq:steady_Kida_vortices} gives us ${ \alpha (\tt_{3}) }$.

\subsection{Models A and B}
\label{sub:Model_AB_models}

We shall first consider two simple prescriptions for ${ \partial_{\tt_{3}} \ln{(\trhoo)} }$: one that corresponds to a constant-mass vortex (\S\ref{ssub:Model_A}), and one that corresponds to incompressible gas (\S\ref{ssub:Model_B}). We will then consider the general case and show that the fixed-mass and incompressible examples neatly isolate the two possible modes of vortex evolution (\S\ref{ssub:Model_C}).

\subsubsection{Model A: fixed-mass vortices}
\label{ssub:Model_A}

\newcommand{\trhogo}{\tilde{\rho}_{g, 0}}

\newcommand{\fdo}{f_{d, 0}}
\newcommand{\fgo}{f_{g, 0}}

Let us first consider the case where ${ \partial_{\tt_{3}} \ln{(\trhoo)} }$ is null. If we interpret ${ \partial_{\tt_{3}} \ln{(\trhoo)} }$ as a proxy for the mass influx at the vortex's boundary, we see that gas must leave the vortex as dust enters. Therefore, the vortex's total mass remains constant.  

In these conditions, we can show sequentially that $\trhoo$, $\TV$, $\tomegao$, $\alpha$, $\tbuo$, $\thK$ and $\tbvi$ are all constant, even on the slow timescale~$\tt_{3}$. Vortex evolution is thus entirely described by\!\!\!\!\!
\begin{subequations}
    \label{eq:Model_A_third_order_densities_equations}
    \begin{align}
        \partial_{\tt_{3}} \ln{(\trhogo)} &= + \fdo \, \tDelta \thK, \label{eq:Model_A_third_order_gas_density_equation} \\
        \partial_{\tt_{3}} \ln{(\trhodo)} &= - \fgo \, \tDelta \thK . \label{eq:Model_A_third_order_dust_density_equation}
    \end{align}
\end{subequations}
Essentially, the dust density increases because the dust drifts towards the pressure maximum at the vortex's centre, but since the vortex's mass is constant, the gas density must decrease in response, so the gas spirals outwards.

At this stage, we should address a possible point of confusion. Conventional wisdom is that slow flows are nearly incompressible, because sound waves quickly erase any density anomaly. But then if vortex evolution is slow, how can the gas density change significantly? To resolve this paradox, consider the Earth's atmosphere: even if it was at rest, its density would be much higher near the surface than near space. This is because sound waves do not nudge the flow towards uniform density, but towards hydrostatic balance. The same thing happens in our system, except that the rotation inherent to vortices means that sound waves nudge the flow towards geostrophic rather than hydrostatic balance. Another useful analogy is with the inward radial drift of solids in \PPDs, which is accompanied by a small outward drift of gas \citep{Nakagawa+1986}.

\newcommand{\tT}{\tilde{T}}

Equations~\eqref{eq:Model_A_third_order_densities_equations} are coupled via the $\fgo$ and $\fdo$ terms on the right hand side. By changing the time variable to ${ \tT = |\tDelta \thK| \, \tt_{3} }$, we can simplify those equations to
\begin{equation}
    \nonumber
    \partial_{\tT} \trhodo = \frac{\trhogo \times \trhodo}{\trhogo + \trhodo} , 
    \quad \text{and} \quad
    \trhogo + \trhodo = \text{C}^{\text{st.}} .
\end{equation}
By choosing the reference density ${ \overline{\rho} = \rhog (t = 0) + \rhod (t = 0) }$, we get ${ \trhogo + \trhodo = 1 }$. The first equation then becomes
\begin{equation}
    \nonumber
    \partial_{\tT} \trhodo = (1 - \trhodo) \, \trhodo .
\end{equation}
We recognise a Bernoulli equation, whose solution is
\begin{subequations}
    \label{eq:Model_A_third_order_densities_solutions}
    \begin{align}
        \trhogo (\tT) &= \frac{\trhogo (0) \, \e^{- \tT}}{\trhodo (0) + \trhogo (0) \, \e^{- \tT}} , \label{eq:Model_A_third_order_gas_density_solution} \\
        \trhodo (\tT) &= \frac{\trhodo (0)}{\trhodo (0) + \trhogo (0) \, \e^{- \tT}}  . \label{eq:Model_A_third_order_dust_density_solution}
    \end{align}
\end{subequations}
In dimensional terms, this translates to
\begin{subequations}
    \label{eq:Model_A}
    \begin{align}
        \rhog (t) &= \frac{\rho (0)}{1 + \dtg (0) \, \e^{\tau \, |\Delta \hK| \, t}} , \label{eq:Model_A_gas_density} \\
        \bug (t) &= \buK , \phantom{\frac{1}{1}} \label{eq:Model_A_gas_velocity} \\
        h (t) &= [1 + \dtg (t)] \, \hK , \phantom{\frac{1}{1}} \label{eq:Model_A_pseudo_pressure} \\
        \rhod (t) &= \frac{\rho (0)}{1 + \left[ \dtg (0) \, \e^{ \tau \, |\Delta \hK| \, t} \right]^{-1}} , \label{eq:Model_A_dust_density} \\
        \bud (t) &= \buK + \tau \bnabla \hK , \label{eq:Model_A_dust_velocity}
    \end{align}
\end{subequations}

Note that in this model, the dust-to-gas ratio grows exponentially, without limit. This result should be taken with a grain of salt, because we assumed that the dust-to-gas ratio is of order unity or less at the start of the derivation.

\subsubsection{Model B: incompressible vortices}
\label{ssub:Model_B}

The previous model absorbed the entirety of the change in specific angular momentum caused by dust compression into gas decompression. Let us now construct a second model that absorbs everything into vorticity. To do so, we assume that the gas is incompressible. Since ${ \bug = \bu - \fd \, \bv }$ and ${ \bnabla \bcdot \bug = 0}$, ${ \tbnabla \bcdot \tbu }$ becomes equal to ${ \fd \, \tbnabla \bcdot \bv }$. Therefore, the mass influx prescription is
\begin{equation}
    \label{eq:Model_0_third_order_mass_influx_prescription}
    \partial_{\tt_{3}} \ln{(\trhoo)} = - \fdo \, \tDelta \thK .
\end{equation}

\newcommand{\APV}{\vartheta}

We can use the conservation of potential vorticity and the fact that $\tomegao$ is only a function of $\alpha$ to show that vortex evolution is then governed by
\begin{subequations}
    \label{eq:Model_B_third_order_equations}
    \begin{align}
        \partial_{\tt{3}} \ln{(\dtgo)} &= - \tDelta \thK , \label{eq:Model_B_third_order_equations_mu} \\
        \partial_{\tt_{3}} \alpha &= \frac{\APV - \TV (\alpha)}{\rd_{\alpha} \TV (\alpha)} \tDelta \thK (\alpha) , \label{eq:Model_B_third_order_equations_alpha}
    \end{align}
\end{subequations}
where ${ \APV = \trhogo \, \PV = \TV / (1 + \dtgo) }$ is an alternative potential vorticity. Eqs.~\eqref{eq:Model_B_third_order_equations} form a triangular system: the second equation is independent of $\dtgo$ and can be solved first.

\newcommand{\omegaK}{\omega_{K}}

But before we do that, let us unpack the physics contained \mbox{in Eqs.~\eqref{eq:Model_B_third_order_equations}. Firstly, Eq.~\eqref{eq:Model_B_third_order_equations_mu} tells us that the dust density} \mbox{increases because the dust spirals towards the vortex's centre.} \mbox{But in doing so it loses angular momentum, so the gas needs} \mbox{to spin up. And since the vortex is fighting against the} Keplerian \mbox{shear, any change in strength causes a change in shape.}

The subtelty is that the dust's specific angular momentum combines vortical rotation and Keplerian rotation. For the anticyclonic vortices of interest, those two contributions take opposite signs. Consequently, the vortex can get stronger or weaker over time depending on the sign of ${ \omegaK + 2 \Omega }$. Specifically, quasi-circular vortices are strong, so they beat the Keplerian term, make the total angular momentum negative, and get stronger over time. Conversely, elongated vortices are weak, so they get weaker over time. For Kida vortices, the boundary between those two regimes is ${ \alpha = 2 + \sqrt{7} \approx 4.6 }$.

Equation~\eqref{eq:Model_B_third_order_equations_alpha} shows two stationary points: ${ \alpha = 2 }$ because ${ \tDelta \thK \! = \! 0 }$ (\textit{cf.} Eq.~\ref{eq:Kida_vortex_P}), and ${ \alpha \! = \! 2 \! + \! \sqrt{7} }$ because ${ \APV \! = \! \PV \! = \! \TV \! = \! 0 }$. However, Eq.~\eqref{eq:Model_AB_third_order_dust_to_gas_ratio} indicates that in this second case, the dust-to-gas ratio increases exponentially with time, so the complete system is not stationary at all. In terms of linear stability, ${ \alpha = 2 }$ is stable and ${ \alpha = 2 + \sqrt{7} }$ unstable.

\begin{figure*}
    \centering
    \vspace{-1 \baselineskip}
    \includegraphics[width = 0.9 \linewidth]{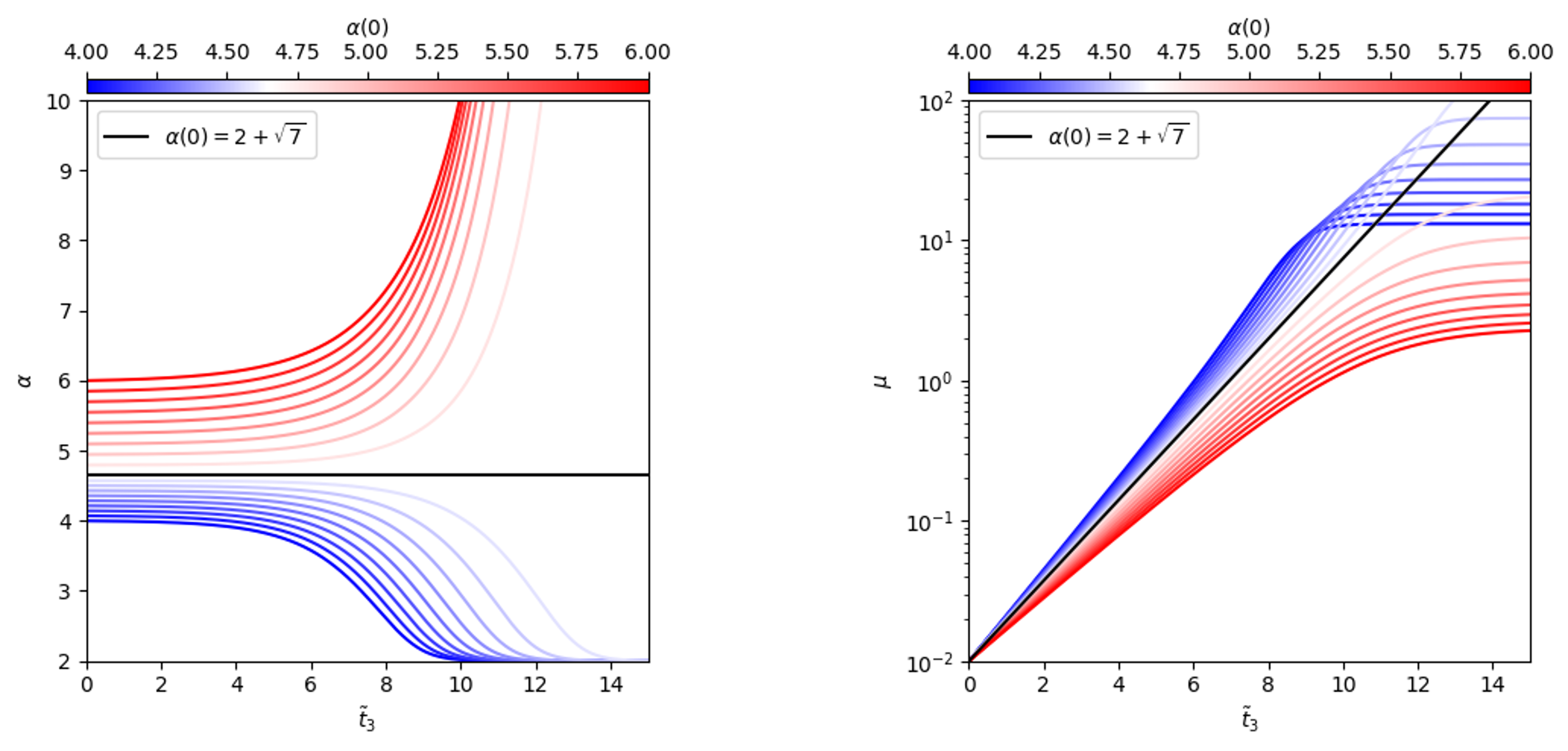}
    \caption{Numerical solution to Eqs.~\eqref{eq:Model_B_third_order_equations}. ${ \alpha }$ is the ratio between the major and minor axes of the vortex, ${ \dtg }$ is the dust-to-gas ratio, and ${ \tt_{3} = \St \times \Omega t }$ is a dimensionless time variable that is convenient when studying dynamics on the dust concentration timescale. We vary the initial aspect ratio ${ \alpha (0) }$, and find two groups of vortices: the red ones are sheared out by the dust, whereas the blue one converge towards the aspect ratio of epicyclic motion, ${ \alpha = 2 }$. Both effectively stop concentrating dust in finite time. \textit{Parameters:} ${ S / \Omega = 3/2 }$, ${ \dtg (0) = 10^{-2} }$.}
    \label{fig:Model_B}
    \vspace{-1 \baselineskip}
\end{figure*}

\phantom{\EI}
\vspace{-1 \baselineskip}

We solve Eqs.~\eqref{eq:Model_B_third_order_equations} numerically in Fig.~\ref{fig:Model_B}. We assumed that the dust-to-gas ratio is initially interstellar, ${ \dtg (0) = 0.01 }$, but we experimented with various initial aspect ratios. We focused our exploration on ${ 4 < \alpha (0) < 6 }$, because outside of this band vortices are subject to the elliptical instability (\EI\ -- \citealt{LesurPapaloizou2009}). It destroys the vortices with~${ \alpha \! < \! 4 }$, and either destroys or makes turbulent those with~${ \alpha \! > \! 6 }$.\footnote{Vortex boundaries may elliptically be unstable even inside the band \citep{LesurPapaloizou2009}, but our focus is on vortex cores.}

The figure confirms that the vortices form two groups, the high-aspect-ratio ‘weak' vortices that get even more elongated over time and the low-aspect-ratio ‘strong’ vortices that get even more circular over time. As expected, the boundary is situated in ${ \alpha (0) = 2 + \sqrt{7} }$.

\newcommand{\tomegaK}{\tilde{\omega}_{K}}

The strong vortices converge to ${ \alpha = 2 }$. This is because dust stops concentrating, thereby halting the spin-up. Conversely, the weak vortices diverge to ${ \alpha = + \infty }$ even though the pressure
Laplacian and therefore the dust concentration rate converge to zero. This is because the residual dust concentration still induces a small spin-up, and ${ \rd_{\alpha} \tomegaK }$ is small when $\alpha$ is large, so this small spin-up results in significant elongation.

Regarding the dust-to-gas ratio, it grows exponentially at first, then saturates. For weak vortices, the dust-to-gas ratio saturates because ${ \Delta \thK }$ scales with ${ 1 / \alpha }$, so once they reach a certain elongation, vortices lose their dust trapping efficiency. For strong vortices, ${ \dtg }$ saturates because ${ \Delta \hK }$ goes to zero when $\alpha$ converges to $2$. When ${ \alpha (0) \approx 2 + \sqrt{7} }$, it takes a long time to reach either saturation point, hence why those vortices exhibit the highest final dust densities.

Interestingly, the final dust-to-gas ratio is independent of the initial dust-to-gas ratio. Indeed, when ${ \dtg (0) \ll 1 }$, ${ \APV }$ becomes a function of ${ \alpha (0) }$ only. Since all strong vortices converge to ${ \TV = - 3.75 }$, conservation of potential vorticity indicates that ${ \dtg_{\infty} \approx - 3.75 / \TV [\alpha (0)] - 1 }$. Similarly, weak vortices all converge to ${ \TV \! = \! 2 }$, so ${ \dtg_{\infty} \! = \! 2 / \TV [\alpha (0)] \! - \! 1 }$. But once again, since we assumed order-unity-or-less dust-to-gas ratios at the start of the derivation, we should be wary of those predictions.

\textit{In fine}, Model `B' boils down to
\begin{subequations}
    \label{eq:Model_B}
    \begin{align}
        \bug (t) &= \buK [\alpha (t) ] , \label{eq:Model_B_gas_velocity} \\
        h (t) &= [1 + \dtg (t)] \, \hK [\alpha (t)] , \label{eq:Model_B_pseudo_pressure} \\
        \bud (t) &= \buK [\alpha (t) ] + \tau \bnabla \hK [\alpha (t)] , \label{eq:Model_B_dust_velocity}
    \end{align}
\end{subequations}
where ${ \alpha (t) }$ and ${ \dtg(t) }$ follow Eqs.~\eqref{eq:Model_B_third_order_equations}. 

\subsubsection{The general case}
\label{ssub:Model_C}

The previous two models rely on very particular prescriptions for the mass influx at the vortex's boundary. To gain in generality, we can write
\begin{equation}
    \label{eq:eq:Model_C_generic_mass_influx}
    \partial_{\tt_{3}} \ln{(\trhoo)} = - \beta \, \fdo \, \tDelta \thK ,
\end{equation}
where ${ \beta }$ is a free parameter. 

Thanks to potential vorticity conservation (Eq.~\ref{eq:conservation_of_potential_vorticity}), the quantitative behavior of the system is easy to predict:
\begin{itemize}
    \vspace{-0.5 \baselineskip}
    \item[(i)] If $\beta$ is positive, $\trhoo$ is increasing so $\TV$ must travel away from zero. So just like in model B, weak vortices become weaker over time and strong vortices become stronger. The boundary between the two regimes is still set by ${\TV = 0}$ and therefore by ${ \alpha = 2 + \sqrt{7} }$.
    \item[(ii)] If $\beta$ is null, we recover model A.
    \item[(iii)] If $\beta$ is negative, $\trhoo$ is decreasing so $\TV$ must travel towards zero. Therefore, all vortices converge to an aspect ratio of ${ 2 + \sqrt{7} }$. In particular, those starting with ${ 4 < \alpha < 6 }$ avoid the \EI\ and amass enormous amounts of dust.
\end{itemize}

Note that while option (iii) seems interesting for planetesimal formation, it requires vortices to lose mass as they gain dust, which seems unlikely. Options (i) and (ii) are more realistic, and are examplified by models~A and~B. This shows the real value of our models: they are simple, pedagogical introductions to the two main modes of dusty vortex evolution.

\section{Discussion}
\label{sec:Discussion}

We have derived two analytical models for the evolution of dust-laden vortex cores in \PPDs. To do so, we relied on a number of strong hypotheses, whose validity we discuss in~\S\ref{sub:Discussion_hypotheses_and_limitations}. As a consequence, our models reflect extreme cases more than astrotypical regimes. Nevertheless, we outline several applications in~\S\ref{sub:Discussion_applications}. Finally, we compare our predictions to numerical experiments in~\S\ref{sub:Discussion_comparison_to_simulations}.

\subsection{Hypotheses and limitations}
\label{sub:Discussion_hypotheses_and_limitations}

\subsubsection{Relevance of the hypotheses}
\label{ssub:Discussion_hypotheses}

We chose to work with elliptical vortices. Numerical experiments suggest that this is a good approximation for large-scale vortices in \PPDs, at least those formed by the \RWI\ \citep{SurvilleBarge2015}.

Furthermore, simulations also support our assumption that the gas density is initially nearly uniform in the vortex's core. Indeed, gas density varies by less than 50 \% between the boundary and the core of vortices formed by the \RWI\ \citep{SurvilleBarge2015} or the \COS\ \citep{Raettig+2021}.

Another questionable point is that all our particles share the same size. This is partially justified by the fact that particles size distribution are narrower in vortices than in the rest of the disc. Indeed, vortices preferentially capture particles of Stokes number ${ \St \approx 1 }$ \citep{BargeSommeria1995}. Furthermore, small particles take a long time to reach the vortex's centre, so vortex cores are also size-sorting.

We specifically work with small solids. One could argue that ${ \St \ll 1 }$ is the relevant regime because this is where fragmentation and bouncing stop collisional growth (see, \textit{e.g.}, \citealt{Drazkowska+2023}). But conversely, vortices preferentially capture particles of Stokes number ${ \St \approx 1 }$, so maybe those matter more. At any rate, the limit of small particles has the benefit of being self-consistent. Indeed, the pressure-less fluid approximation is only accurate when ${ \St \ll 1 }$ \citep{Garaud+2004}.

Finally and contrary to \cite{LyraLin2013}, we neglect dust diffusion. This may be incorrect if there is some small-scale turbulence inside the vortex. However, whether vortex cores are turbulent is an open question, and even if they are it is hard to determine how much diffusion this turbulence induces. So to cover all bases, we should also study vortex evolution in the laminar case. In that sense, our paper and theirs are complementary.

\subsubsection{Other limitations}
\label{ssub:Discussion_limitations}

Our vortex models are 2D. This assumption is motivated by the Taylor-Proudman effect, which favours columnar structures. In particular, \cite{RailtonPapaloizou2014} show that in local and vertically isothermal models, the Navier-Stokes equations admit exact columnar vortex solutions. However, in more complete models, this is not necessarily true. For instance, \cite{Richard+2013} find that the midplane gas shoots up towards the disc’s surfaces. These upflows could entrain some dust and reduce the midplane dust-to-gas ratio.

In the same vein, we do not model the vortex boundary. Presumably, dust is delivered to the vortex by the radial drift, so it enters at the leading outer edge. Consequently, it is not uniformly distributed inside the vortex: it only occupies a spiral stemming from the entry point. Furthermore, to maintain uniformity, our models requires an exponentially growing influx of dust. In reality, the dust influx is probably nearly constant, leading to larger dust densities in the centre than near the boundary. As for the gas influx, it is unlikely to take just the right value to compensate the dust influx (model A), or to be exactly null (model B). So let us stress once again that our models are not quantitative predictions. Their goal is only to illustrate the two main modes of vortex response to dust capture.

Another issue related to the vortex's boundary is the instability discovered numerically by \cite{LesurPapaloizou2009}. Indeed, it could destroy the vortex and/or affect the exchanges of dust between the disc and the vortex's core. \cite{Dritschel1990} also discovered an instability affecting the boundary of elliptical vortices, but we do not know if this instability is active in accretion discs. At any rate, those instabilities would only reduce the amount of time available to concentrate dust, so our estimates for the final dust-to-gas ratio are upper bounds.

Model A also breaks down in the Epstein regime, because the gas density changes, so $\St$ becomes time-dependent and cannot be used as a book-keeping parameter anymore. That being said, when the dust-to-gas ratio is much smaller than one, $\St$ only varies by percentage points, not orders of magnitude. Our method would therefore be easy to extend, except that the solution would switch from analytical to semi-analytical. But we do not expect to see anything fundamentally new. We only expect larger relative velocities between dust and gas, and therefore faster vortex evolution.

Ultimately, the main limitation is that we implicitly assume in \S\ref{ssub:Model_AB_second_order_centre_of_mass_velocity} that there is no instability affecting vortex evolution on the orbital timescale. We adress this issue in paper II.

\subsection{Applications}
\label{sub:Discussion_applications}

\subsubsection{Providing intuition}
\label{ssub:Discussion_applications_intuition}

The capture of test particles inside vortices had been studied before \citep{BargeSommeria1995, AdamsWatkins1995, Tanga+1996, Chavanis2000}, but always with a Lagrangian view. Our Eulerian approach makes it evident that there is a strong analogy between the dust spiral towards vortex centres and the radial drift in discs.

The regime of inertial particles had almost never been studied before, except by \cite{Surville+2016} with their zero-dimensional model. Our models rely on extreme mass influx prescriptions, so they are not intended to make quantitative predictions of protoplanetary vortex evolution. Their goal is rather to isolate the main effect of dust concentration (it modifies the dust's specific angular momentum), and the two possible gas responses (either its density decreases, or the vortex strength and shape change). We show in \S\ref{ssub:Model_C} that real protoplanetary vortices combine both mechanisms, with relative contributions determined by the boundary dynamics.

\subsubsection{Applications to planet formation theory}
\label{ssub:Discussion_applications_to_planet_formation_theory}

We stated in \S\ref{ssub:Model_A} and \S\ref{ssub:Model_B} that since we assumed that the dust-to-gas ratio is of order unity or less at the start of the our derivation, we cannot trust the predictions our models make regarding the existence of a maximal dust-to-gas ratio. However, a more precise statement is that if we let~$\dtg$ be big, the convergence of our approximation stops being uniform in~$\St$. Rather, the Stokes number at which our approximation becomes accurate becomes smaller and smaller as $\dtg$ increases. 

\begin{figure}
    \centering
    \includegraphics[width = \linewidth]{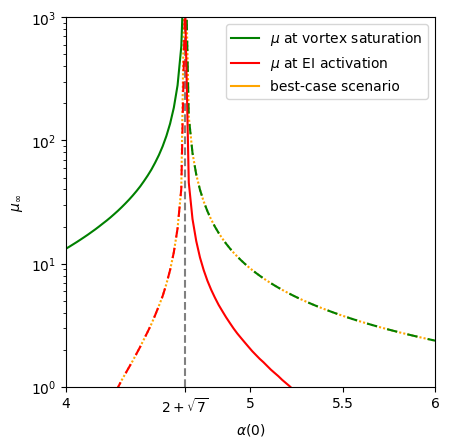}
    \caption{Maximal dust-to-gas ratio reached by vortices as a function of their initial aspect ratio, according to model B. The green line lets the vortices evolve forever whereas the red line stops the simulation when ${ \alpha = 4 }$ or ${ \alpha = 6 }$. This is an attempt at representing the effect of the \EI. Finally, the orange line stops the simulation when ${ \alpha = 4 }$, in order to represent a best-case scenario where the high-aspect-ratio branch of the \EI\ does not affect vortex evolution. \textit{Parameters:} ${ S / \Omega = 3/2 }$, ${ \dtg (0) = 10^{-2} }$.}
    \label{fig:maximal_dtg}
\end{figure}

Model B predicts a maximal dust-to-gas ratio, so if the particles are small enough, it remains accurate at all times. In that case, the green curve of Fig.~\ref{fig:maximal_dtg} shows that the maximal dust-to-gas ratio is rarely larger than 100. This is the typical threshold for gravitational collapse, suggesting that laminar concentration is not sufficient to create planetesimals.

Futhermore, Fig.~\ref{fig:Model_B}-\textit{left} shows that all vortices enter an elliptically unstable band well before reaching their maximal dust-to-gas ratio. If a vortex has a low aspect ratio, the \EI\ destroys it \citep{LesurPapaloizou2009}. If the vortex has a high aspect ratio, it may survive, but becomes turbulent. This may induce dust diffusion, which would balance inward migration and prevent further migration \citep{LyraLin2013}. The red curve of Fig.~\ref{fig:maximal_dtg} shows that this limits the dust-to-gas ratio even more than saturation.

Now since dust diffusion is a controversial idea, we show with the orange curve of Fig.~\ref{fig:maximal_dtg} a best-case-scenario where low-aspect-ratio vortices are destroyed by the \EI\ while high-aspect-ratio vortices are completely unaffected. Even then, few vortices reach a dust-to-gas ratio of 10, let alone 100.

We think this makes a convincing case againts the laminar vortex pathway to planetesimal formation.

\subsubsection{Vortex observations (at the population level)}
\label{ssub:Discussion_vortex_observations}

Model B predicts that ${ \alpha = 2 + \sqrt{7} }$ is an unstable equilibrium, expelling nearly all vortices to the elliptically unstable bands of \cite{LesurPapaloizou2009} in ten vortex evolution timescales ${ \Omega / (\St \, | \Delta \hK |) }$ or less. While the tipping point’s exact value is model-dependent, the idea that dust brings all vortices to the elliptically unstable bands seems robust. And if the high-$\alpha$ branch of the \EI\ destroys vortices, this unstable equilibrium has two notable consequences.

Firstly, it sets an upper bound on vortex lifetimes. Indeed, we expect all vortices filled with small dust particles to be destroyed in a few slow timescales by the \EI. This ‘vortex life expectancy’ could be combined with the fraction of discs observed to host a vortex to constrain the vortex creation frequency. One difficulty is that the vortex life expectancy depends on ${ \St }$, so grain size will need to be controlled for.

Secondly, core-stretching affects the distribution of vortex aspect ratios. Indeed, Fig.~\ref{fig:Model_B} suggests a double-peaked distribution, with peaks at the boundaries of the elliptically unstable bands. In the future, this prediction could be tested observationally by studying vortices at the population level.

\subsection{Comparison to previous models}
\label{sub:Discussion_comparison_to_simulations}

As demonstrated in \S\ref{sub:Model_0_model}, model 0 recover the dust concentration timescale of \cite{Chavanis2000}. Our prediction for the $e$-folding rate of the dust-to-gas ratio is also similar to that of \cite{Surville+2016}, even though their model includes the disc's large-scale pressure gradient.

\cite{Fu+2014} and \cite{Surville+2016} show that the dust forms a spiral inside the vortex. This is consistent with our prediction for the trajectory of individual particles, but not with our assumption that the dust-to-gas ratio is uniform. This discrepancy is certainly important in the outer layers of vortices, but the spiral is tighltly wound so any amount of diffusion would make the vortex core uniform. This may be what happens in figure~1e of \cite{Fu+2014}.

Model B predicts that weak vortices become weaker over time. This `core-weakening' effect was reported but left unexplained by \cite{CrnkovicRubsamen+2015}, \cite{Surville+2016} and \cite{Miranda+2017}. \cite{Fu+2014} also show it in their figure~3a, but do not discuss it. The only caveat is that \cite{CrnkovicRubsamen+2015} reports core weakening even for strong vortices, and only for large particles.

Finally, model B predicts that weak vortices become more and more elongated over time. This `core-stretching' effect was observed by \cite{CrnkovicRubsamen+2015} and \cite{Miranda+2017}. \cite{Fu+2014} and \cite{Surville+2016} also show it some of their figures, but do not comment on it. The same caveat as for core weakening applies.

\section{Conclusion}
\label{sec:Conclusion}

This series is concerned with what happens when one adds dust to a vortex embedded in a \PPD. In the present paper, we first reproduce the well-known result that if a protoplanetary vortex is weak and anticyclonic, then its centre is a pressure maximum and the dust spirals towards it. We then study how dust concentration affects the vortex' long term evolution. 

We find that as dust gets closer to the vortex's centre, its angular momentum changes. Total angular momentum is conserved, so the gas must respond in one of two ways: either it moves away from the vortex's centre, or it adapts its vorticity. If vorticity decreases, then the Keplerian shear flow surrounding the vortex makes it more elongated. Those `core weakening' and `core stretching' effects had already been seen in simulations \citep{Fu+2014, CrnkovicRubsamen+2015, Surville+2016, Miranda+2017}. 

Conversely, if vorticity increases, the vortex becomes more circular. Either way, all vortices eventually reach an elliptically unstable state. And since the \EI\ can destroy vortices, dust-induced deformation may set the life expectancy of vortices, and may induce observable bimodality in the distribution of vortex aspect ratios.

But more importantly, it limits the time available for vortices to concentrate dust. We find that the dust-to-gas ratio at the time of \EI\ activation is almost always below the threshold for gravitational collapse. Furthermore, this result remains true even when we assume the high-aspect-ratio band of the \EI\ does not affect vortex evolution. Indeed, dust makes weak vortices weaker, so they eventually stop concentrating dust. We call this process `saturation'. We plug in the numbers in Fig.~\ref{fig:maximal_dtg} and conclude that the `laminar' vortex pathway to planetesimal formation is not viable.

One last application of our models will be highlighted in paper II: they provide the background flow for a linear stability analysis of dusty vortices. This will allow us to show that the streaming instability remains active in vortices -- a major step towards validating the `turbulent' vortex pathway to planetesimal formation.

Now because our work is analytical, we had to make many assumptions. We represented dust as a pressure-less fluid, we assumed that all the particles have the same size, we limited ourselves to the regime of dilute and small dust, we neglected viscosity and turbulent diffusion, we worked in 2D, and we did not model what happens at the vortex’s boundary. But arguably, our most limiting assumption is that gas and dust density remain uniform at all times. This is mathematically possible for Kida vortices, but physically implausible.

\section*{Acknowledgments}

We wish to thank the anonymous referee and Andrew Youdin for their advice that helped us clarify several parts of the manuscript. Support for N.M. was provided by a Cambridge International \& Isaac Newton Studentship.

\section*{Data Availability}

The data and numerical codes underlying this article were produced by the authors. They will be shared on reasonable request to the corresponding author.

\bibliographystyle{mnras}
\bibliography{main}

@ARTICLE{GoldreichLyndenBell1965,
       author = {{Goldreich}, P. and {Lynden-Bell}, D.},
        title = "{II. Spiral arms as sheared gravitational instabilities}",
      journal = {\mnras},
         year = 1965,
        month = jan,
       volume = {130},
        pages = {125},
          doi = {10.1093/mnras/130.2.125},
       adsurl = {https://ui.adsabs.harvard.edu/abs/1965MNRAS.130..125G}
}

@ARTICLE{Hawley1995,
       author = {{Hawley}, John F. and {Gammie}, Charles F. and {Balbus}, Steven A.},
        title = "{Local Three-dimensional Magnetohydrodynamic Simulations of Accretion Disks}",
      journal = {\apj},
         year = 1995,
        month = feb,
       volume = {440},
        pages = {742},
          doi = {10.1086/175311},
       adsurl = {https://ui.adsabs.harvard.edu/abs/1995ApJ...440..742H}
}

@ARTICLE{LatterPapaloizou2017,
       author = {{Latter}, Henrik N. and {Papaloizou}, John},
        title = "{Local models of astrophysical discs}",
      journal = {\mnras},
         year = 2017,
        month = dec,
       volume = {472},
       number = {2},
        pages = {1432-1446},
          doi = {10.1093/mnras/stx2038},
archivePrefix = {arXiv},
       eprint = {1708.09285},
 primaryClass = {astro-ph.HE},
       adsurl = {https://ui.adsabs.harvard.edu/abs/2017MNRAS.472.1432L}
}

@ARTICLE{YoudinGoodman2005,
       author = {{Youdin}, Andrew N. and {Goodman}, Jeremy},
        title = "{Streaming Instabilities in Protoplanetary Disks}",
      journal = {\apj},
         year = 2005,
        month = feb,
       volume = {620},
       number = {1},
        pages = {459-469},
          doi = {10.1086/426895},
archivePrefix = {arXiv},
       eprint = {astro-ph/0409263},
 primaryClass = {astro-ph},
       adsurl = {https://ui.adsabs.harvard.edu/abs/2005ApJ...620..459Y}
}

@ARTICLE{LesurPapaloizou2009,
       author = {{Lesur}, G. and {Papaloizou}, J.~C.~B.},
        title = "{On the stability of elliptical vortices in accretion discs}",
      journal = {\aap},
         year = 2009,
        month = apr,
       volume = {498},
       number = {1},
        pages = {1-12},
          doi = {10.1051/0004-6361/200811577},
archivePrefix = {arXiv},
       eprint = {0903.1720},
 primaryClass = {astro-ph.EP},
       adsurl = {https://ui.adsabs.harvard.edu/abs/2009A&A...498....1L}
}

@ARTICLE{Chavanis2000,
       author = {{Chavanis}, P.~H.},
        title = "{Trapping of dust by coherent vortices in the solar nebula}",
      journal = {\aap},
         year = 2000,
        month = apr,
       volume = {356},
        pages = {1089-1111},
archivePrefix = {arXiv},
       eprint = {astro-ph/9912087},
 primaryClass = {astro-ph},
       adsurl = {https://ui.adsabs.harvard.edu/abs/2000A&A...356.1089C}
}

@ARTICLE{Kida1981,
      author = {Kida ,Shigeo},
      title = {Motion of an Elliptic Vortex in a Uniform Shear Flow},
      journal = {Journal of the Physical Society of Japan},
      volume = {50},
      number = {10},
      pages = {3517-3520},
      year = {1981},
      doi = {10.1143/JPSJ.50.3517},
      URL = {https://doi.org/10.1143/JPSJ.50.3517},
      eprint = {https://doi.org/10.1143/JPSJ.50.3517}
}

@ARTICLE{RailtonPapaloizou2014,
       author = {{Railton}, A.~D. and {Papaloizou}, J.~C.~B.},
        title = "{On the local stability of vortices in differentially rotating discs}",
      journal = {\mnras},
         year = 2014,
        month = dec,
       volume = {445},
       number = {4},
        pages = {4409-4426},
          doi = {10.1093/mnras/stu2060},
archivePrefix = {arXiv},
       eprint = {1410.1323},
 primaryClass = {astro-ph.EP},
       adsurl = {https://ui.adsabs.harvard.edu/abs/2014MNRAS.445.4409R}
}

@ARTICLE{SurvilleBarge2015,
       author = {{Surville}, Cl{\'e}ment and {Barge}, Pierre},
        title = "{Quasi-steady vortices in protoplanetary disks. I. From dwarfs to giants}",
      journal = {\aap},
         year = 2015,
        month = jul,
       volume = {579},
          eid = {A100},
        pages = {A100},
          doi = {10.1051/0004-6361/201424663},
       adsurl = {https://ui.adsabs.harvard.edu/abs/2015A&A...579A.100S}
}

@ARTICLE{Raettig+2021,
       author = {{Raettig}, Natalie and {Lyra}, Wladimir and {Klahr}, Hubert},
        title = "{Pebble Trapping in Vortices: Three-dimensional Simulations}",
      journal = {\apj},
         year = 2021,
        month = jun,
       volume = {913},
       number = {2},
          eid = {92},
        pages = {92},
          doi = {10.3847/1538-4357/abf739},
archivePrefix = {arXiv},
       eprint = {2103.04476},
 primaryClass = {astro-ph.EP},
       adsurl = {https://ui.adsabs.harvard.edu/abs/2021ApJ...913...92R}
}

@INPROCEEDINGS{Drazkowska+2023,
       author = {{Drazkowska}, J. and {Bitsch}, B. and {Lambrechts}, M. and {Mulders}, G. D. and {Harsono}, D. and {Vazan}, A. and {Liu}, B. and {Ormel}, C. W. and {Kretke}, K. and {Morbidelli}, A.},
        title = "{Planet Formation Theory in the Era of ALMA and Kepler: from Pebbles to Exoplanets}",
    booktitle = {Protostars and Planets VII},
         year = 2023,
       editor = {{Inutsuka}, S. and {Aikawa}, Y. and {Muto}, T. and {Tomida}, K. and {Tamura}, M.},
        pages = {717},
       adsurl = {https://ui.adsabs.harvard.edu/abs/2023ASPC..534..717D}
}

@ARTICLE{BargeSommeria1995,
       author = {{Barge}, P. and {Sommeria}, J.},
        title = "{Did planet formation begin inside persistent gaseous vortices?}",
      journal = {\aap},
         year = 1995,
        month = mar,
       volume = {295},
        pages = {L1-L4},
          doi = {10.48550/arXiv.astro-ph/9501050},
archivePrefix = {arXiv},
       eprint = {astro-ph/9501050},
 primaryClass = {astro-ph},
       adsurl = {https://ui.adsabs.harvard.edu/abs/1995A&A...295L...1B}
}

@ARTICLE{LyraLin2013,
       author = {{Lyra}, Wladimir and {Lin}, Min-Kai},
        title = "{Steady State Dust Distributions in Disk Vortices: Observational Predictions and Applications to Transitional Disks}",
      journal = {\apj},
         year = 2013,
        month = sep,
       volume = {775},
       number = {1},
          eid = {17},
        pages = {17},
          doi = {10.1088/0004-637X/775/1/17},
archivePrefix = {arXiv},
       eprint = {1307.3770},
 primaryClass = {astro-ph.EP},
       adsurl = {https://ui.adsabs.harvard.edu/abs/2013ApJ...775...17L}
}

@ARTICLE{Richard+2013,
       author = {{Richard}, S. and {Barge}, P. and {Le Diz{\`e}s}, S.},
        title = "{Structure, stability, and evolution of 3D Rossby vortices in protoplanetary disks}",
      journal = {\aap},
         year = 2013,
        month = nov,
       volume = {559},
          eid = {A30},
        pages = {A30},
          doi = {10.1051/0004-6361/201322175},
archivePrefix = {arXiv},
       eprint = {1309.3486},
 primaryClass = {astro-ph.EP},
       adsurl = {https://ui.adsabs.harvard.edu/abs/2013A&A...559A..30R}
}

@ARTICLE{AdamsWatkins1995,
       author = {{Adams}, Fred C. and {Watkins}, Richard},
        title = "{Vortices in Circumstellar Disks}",
      journal = {\apj},
         year = 1995,
        month = sep,
       volume = {451},
        pages = {314},
          doi = {10.1086/176221},
archivePrefix = {arXiv},
       eprint = {astro-ph/9501039},
 primaryClass = {astro-ph},
       adsurl = {https://ui.adsabs.harvard.edu/abs/1995ApJ...451..314A}
}

@ARTICLE{Tanga+1996,
       author = {{Tanga}, P. and {Babiano}, A. and {Dubrulle}, B. and {Provenzale}, A.},
        title = "{Forming Planetesimals in Vortices}",
      journal = {\icarus},
         year = 1996,
        month = may,
       volume = {121},
       number = {1},
        pages = {158-170},
          doi = {10.1006/icar.1996.0076},
       adsurl = {https://ui.adsabs.harvard.edu/abs/1996Icar..121..158T}
}

@ARTICLE{Fu+2014,
       author = {{Fu}, Wen and {Li}, Hui and {Lubow}, Stephen and {Li}, Shengtai and {Liang}, Edison},
        title = "{Effects of Dust Feedback on Vortices in Protoplanetary Disks}",
      journal = {\apjl},
         year = 2014,
        month = nov,
       volume = {795},
       number = {2},
          eid = {L39},
        pages = {L39},
          doi = {10.1088/2041-8205/795/2/L39},
archivePrefix = {arXiv},
       eprint = {1410.4196},
 primaryClass = {astro-ph.EP},
       adsurl = {https://ui.adsabs.harvard.edu/abs/2014ApJ...795L..39F}
}

@ARTICLE{CrnkovicRubsamen+2015,
       author = {{Crnkovic-Rubsamen}, Ivo and {Zhu}, Zhaohuan and {Stone}, James M.},
        title = "{Survival and structure of dusty vortices in protoplanetary discs}",
      journal = {\mnras},
         year = 2015,
        month = jul,
       volume = {450},
       number = {4},
        pages = {4285-4291},
          doi = {10.1093/mnras/stv828},
       adsurl = {https://ui.adsabs.harvard.edu/abs/2015MNRAS.450.4285C}
}

@ARTICLE{Surville+2016,
       author = {{Surville}, Cl{\'e}ment and {Mayer}, Lucio and {Lin}, Douglas N.~C.},
        title = "{Dust Capture and Long-lived Density Enhancements Triggered by Vortices in 2D Protoplanetary Disks}",
      journal = {\apj},
         year = 2016,
        month = nov,
       volume = {831},
       number = {1},
          eid = {82},
        pages = {82},
          doi = {10.3847/0004-637X/831/1/82},
archivePrefix = {arXiv},
       eprint = {1601.05945},
 primaryClass = {astro-ph.EP},
       adsurl = {https://ui.adsabs.harvard.edu/abs/2016ApJ...831...82S}
}

@ARTICLE{Miranda+2017,
       author = {{Miranda}, Ryan and {Li}, Hui and {Li}, Shengtai and {Jin}, Sheng},
        title = "{Long-lived Dust Asymmetries at Dead Zone Edges in Protoplanetary Disks}",
      journal = {\apj},
         year = 2017,
        month = feb,
       volume = {835},
       number = {2},
          eid = {118},
        pages = {118},
          doi = {10.3847/1538-4357/835/2/118},
archivePrefix = {arXiv},
       eprint = {1610.01977},
 primaryClass = {astro-ph.EP},
       adsurl = {https://ui.adsabs.harvard.edu/abs/2017ApJ...835..118M}
}

@INPROCEEDINGS{Bae+2023,
       author = {{Bae}, J. and {Isella}, A. and {Zhu}, Z. and {Martin}, R. and {Okuzumi}, S. and {Suriano}, S.},
        title = "{Structured Distributions of Gas and Solids in Protoplanetary Disks}",
    booktitle = {Protostars and Planets VII},
         year = 2023,
       editor = {{Inutsuka}, S. and {Aikawa}, Y. and {Muto}, T. and {Tomida}, K. and {Tamura}, M.},
        pages = {423}
}

@ARTICLE{Varga+2021,
       author = {{Varga}, J. and {Hogerheijde}, M. and {van Boekel}, R. and {Klarmann}, L. and {Petrov}, R. and {Waters}, L.~B.~F.~M. and {Lagarde}, S. and {Pantin}, E. and {Berio}, Ph. and {Weigelt}, G. and {Robbe-Dubois}, S. and {Lopez}, B. and {Millour}, F. and {Augereau}, J. -C. and {Meheut}, H. and {Meilland}, A. and {Henning}, Th. and {Jaffe}, W. and {Bettonvil}, F. and {Bristow}, P. and {Hofmann}, K. -H. and {Matter}, A. and {Zins}, G. and {Wolf}, S. and {Allouche}, F. and {Donnan}, F. and {Schertl}, D. and {Dominik}, C. and {Heininger}, M. and {Lehmitz}, M. and {Cruzal{\`e}bes}, P. and {Glindemann}, A. and {Meisenheimer}, K. and {Paladini}, C. and {Sch{\"o}ller}, M. and {Woillez}, J. and {Venema}, L. and {Kokoulina}, E. and {Yoffe}, G. and {{\'A}brah{\'a}m}, P. and {Abadie}, S. and {Abuter}, R. and {Accardo}, M. and {Adler}, T. and {Ag{\'o}cs}, T. and {Antonelli}, P. and {B{\"o}hm}, A. and {Bailet}, C. and {Bazin}, G. and {Beckmann}, U. and {Beltran}, J. and {Boland}, W. and {Bourget}, P. and {Brast}, R. and {Bresson}, Y. and {Burtscher}, L. and {Castillo}, R. and {Chelli}, A. and {Cid}, C. and {Clausse}, J. -M. and {Connot}, C. and {Conzelmann}, R.~D. and {Danchi}, W. -C. and {De Haan}, M. and {Delbo}, M. and {Ebert}, M. and {Elswijk}, E. and {Fantei}, Y. and {Frahm}, R. and {G{\'a}mez Rosas}, V. and {Gabasch}, A. and {Gallenne}, A. and {Garces}, E. and {Girard}, P. and {Gont{\'e}}, F.~Y.~J. and {Gonz{\'a}lez Herrera}, J.~C. and {Graser}, U. and {Guajardo}, P. and {Guitton}, F. and {Haubois}, X. and {Hron}, J. and {Hubin}, N. and {Huerta}, R. and {Isbell}, J.~W. and {Ives}, D. and {Jakob}, G. and {Jask{\'o}}, A. and {Jochum}, L. and {Klein}, R. and {Kragt}, J. and {Kroes}, G. and {Kuindersma}, S. and {Labadie}, L. and {Laun}, W. and {Le Poole}, R. and {Leinert}, C. and {Lizon}, J. -L. and {Lopez}, M. and {M{\'e}rand}, A. and {Marcotto}, A. and {Mauclert}, N. and {Maurer}, T. and {Mehrgan}, L.~H. and {Meisner}, J. and {Meixner}, K. and {Mellein}, M. and {Mohr}, L. and {Morel}, S. and {Mosoni}, L. and {Navarro}, R. and {Neumann}, U. and {Nu{\ss}baum}, E. and {Pallanca}, L. and {Pasquini}, L. and {Percheron}, I. and {Pott}, J. -U. and {Pozna}, E. and {Ridinger}, A. and {Rigal}, F. and {Riquelme}, M. and {Rivinius}, Th. and {Roelfsema}, R. and {Rohloff}, R. -R. and {Rousseau}, S. and {Schuhler}, N. and {Schuil}, M. and {Soulain}, A. and {Stee}, P. and {Stephan}, C. and {ter Horst}, R. and {Tromp}, N. and {Vakili}, F. and {van Duin}, A. and {Vinther}, J. and {Wittkowski}, M. and {Wrhel}, F.},
        title = "{The asymmetric inner disk of the Herbig Ae star HD 163296 in the eyes of VLTI/MATISSE: evidence for a vortex?}",
      journal = {\aap},
         year = 2021,
        month = mar,
       volume = {647},
          eid = {A56},
        pages = {A56},
          doi = {10.1051/0004-6361/202039400},
archivePrefix = {arXiv},
       eprint = {2012.05697},
 primaryClass = {astro-ph.SR},
       adsurl = {https://ui.adsabs.harvard.edu/abs/2021A&A...647A..56V}
}

@ARTICLE{Gravity2021,
       author = {{GRAVITY collaboration} and {Sanchez-Bermudez}, J. and {Caratti O Garatti}, A. and {Garcia Lopez}, R. and {Perraut}, K. and {Labadie}, L. and {Benisty}, M. and {Brandner}, W. and {Dougados}, C. and {Garcia}, P.~J.~V. and {Henning}, Th. and {Klarmann}, L. and {Amorim}, A. and {Baub{\"o}ck}, M. and {Berger}, J.~P. and {Le Bouquin}, J.~B. and {Caselli}, P. and {Cl{\'e}net}, Y. and {Coud{\'e} Du Foresto}, V. and {de Zeeuw}, P.~T. and {Drescher}, A. and {Duvert}, G. and {Eckart}, A. and {Eisenhauer}, F. and {Filho}, M. and {Gao}, F. and {Gendron}, E. and {Genzel}, R. and {Gillessen}, S. and {Grellmann}, R. and {Heissel}, G. and {Horrobin}, M. and {Hubert}, Z. and {Jim{\'e}nez-Rosales}, A. and {Jocou}, L. and {Kervella}, P. and {Lacour}, S. and {Lapeyr{\`e}re}, V. and {L{\'e}na}, P. and {Ott}, T. and {Paumard}, T. and {Perrin}, G. and {Pineda}, J.~E. and {Rodr{\'\i}guez-Coira}, G. and {Rousset}, G. and {Segura-Cox}, D.~M. and {Shangguan}, J. and {Shimizu}, T. and {Stadler}, J. and {Straub}, O. and {Straubmeier}, C. and {Sturm}, E. and {van Dishoeck}, E. and {Vincent}, F. and {von Fellenberg}, S.~D. and {Widmann}, F. and {Woillez}, J.},
        title = "{The GRAVITY young stellar object survey. VI. Mapping the variable inner disk of HD 163296 at sub-au scales}",
      journal = {\aap},
         year = 2021,
        month = oct,
       volume = {654},
          eid = {A97},
        pages = {A97},
          doi = {10.1051/0004-6361/202039600},
archivePrefix = {arXiv},
       eprint = {2107.02391},
 primaryClass = {astro-ph.SR},
       adsurl = {https://ui.adsabs.harvard.edu/abs/2021A&A...654A..97G}
}

@ARTICLE{Ragusa+2017,
       author = {{Ragusa}, Enrico and {Dipierro}, Giovanni and {Lodato}, Giuseppe and {Laibe}, Guillaume and {Price}, Daniel J.},
        title = "{On the origin of horseshoes in transitional discs}",
      journal = {\mnras},
         year = 2017,
        month = jan,
       volume = {464},
       number = {2},
        pages = {1449-1455},
          doi = {10.1093/mnras/stw2456},
archivePrefix = {arXiv},
       eprint = {1609.08159},
 primaryClass = {astro-ph.EP},
       adsurl = {https://ui.adsabs.harvard.edu/abs/2017MNRAS.464.1449R}
}

@ARTICLE{Long+2022,
       author = {{Long}, Feng and {Andrews}, Sean M. and {Zhang}, Shangjia and {Qi}, Chunhua and {Benisty}, Myriam and {Facchini}, Stefano and {Isella}, Andrea and {Wilner}, David J. and {Bae}, Jaehan and {Huang}, Jane and {Loomis}, Ryan A. and {{\"O}berg}, Karin I. and {Zhu}, Zhaohuan},
        title = "{ALMA Detection of Dust Trapping around Lagrangian Points in the LkCa 15 Disk}",
      journal = {\apjl},
         year = 2022,
        month = sep,
       volume = {937},
       number = {1},
          eid = {L1},
        pages = {L1},
          doi = {10.3847/2041-8213/ac8b10},
archivePrefix = {arXiv},
       eprint = {2209.05535},
 primaryClass = {astro-ph.EP},
       adsurl = {https://ui.adsabs.harvard.edu/abs/2022ApJ...937L...1L}
}

@ARTICLE{Price+2018,
       author = {{Price}, Daniel J. and {Cuello}, Nicol{\'a}s and {Pinte}, Christophe and {Mentiplay}, Daniel and {Casassus}, Simon and {Christiaens}, Valentin and {Kennedy}, Grant M. and {Cuadra}, Jorge and {Sebastian Perez}, M. and {Marino}, Sebastian and {Armitage}, Philip J. and {Zurlo}, Alice and {Juhasz}, Attila and {Ragusa}, Enrico and {Laibe}, Guillaume and {Lodato}, Giuseppe},
        title = "{Circumbinary, not transitional: on the spiral arms, cavity, shadows, fast radial flows, streamers, and horseshoe in the HD 142527 disc}",
      journal = {\mnras},
         year = 2018,
        month = jun,
       volume = {477},
       number = {1},
        pages = {1270-1284},
          doi = {10.1093/mnras/sty647},
archivePrefix = {arXiv},
       eprint = {1803.02484},
 primaryClass = {astro-ph.SR},
       adsurl = {https://ui.adsabs.harvard.edu/abs/2018MNRAS.477.1270P}
}

@ARTICLE{Ribas+2024,
       author = {{Ribas}, {\'A}lvaro and {Clarke}, Cathie J. and {Zagaria}, Francesco},
        title = "{Inner walls or vortices? Crescent-shaped asymmetries in ALMA observations of protoplanetary discs}",
      journal = {arXiv e-prints},
         year = 2024,
        month = jun,
          eid = {arXiv:2406.14626},
        pages = {arXiv:2406.14626},
          doi = {10.48550/arXiv.2406.14626},
archivePrefix = {arXiv},
       eprint = {2406.14626},
 primaryClass = {astro-ph.EP},
       adsurl = {https://ui.adsabs.harvard.edu/abs/2024arXiv240614626R}
}

@ARTICLE{Lovelace+1999,
       author = {{Lovelace}, R.~V.~E. and {Li}, H. and {Colgate}, S.~A. and {Nelson}, A.~F.},
        title = "{Rossby Wave Instability of Keplerian Accretion Disks}",
      journal = {\apj},
         year = 1999,
        month = mar,
       volume = {513},
       number = {2},
        pages = {805-810},
          doi = {10.1086/306900},
archivePrefix = {arXiv},
       eprint = {astro-ph/9809321},
 primaryClass = {astro-ph},
       adsurl = {https://ui.adsabs.harvard.edu/abs/1999ApJ...513..805L}
}

@ARTICLE{Li+2000,
       author = {{Li}, H. and {Finn}, J.~M. and {Lovelace}, R.~V.~E. and {Colgate}, S.~A.},
        title = "{Rossby Wave Instability of Thin Accretion Disks. II. Detailed Linear Theory}",
      journal = {\apj},
         year = 2000,
        month = apr,
       volume = {533},
       number = {2},
        pages = {1023-1034},
          doi = {10.1086/308693},
archivePrefix = {arXiv},
       eprint = {astro-ph/9907279},
 primaryClass = {astro-ph},
       adsurl = {https://ui.adsabs.harvard.edu/abs/2000ApJ...533.1023L}
}

@ARTICLE{KlahrBodenheimer2003,
       author = {{Klahr}, H.~H. and {Bodenheimer}, P.},
        title = "{Turbulence in Accretion Disks: Vorticity Generation and Angular Momentum Transport via the Global Baroclinic Instability}",
      journal = {\apj},
         year = 2003,
        month = jan,
       volume = {582},
       number = {2},
        pages = {869-892},
          doi = {10.1086/344743},
archivePrefix = {arXiv},
       eprint = {astro-ph/0211629},
 primaryClass = {astro-ph},
       adsurl = {https://ui.adsabs.harvard.edu/abs/2003ApJ...582..869K}
}

@ARTICLE{Petersen+2007,
       author = {{Petersen}, Mark R. and {Julien}, Keith and {Stewart}, Glen R.},
        title = "{Baroclinic Vorticity Production in Protoplanetary Disks. I. Vortex Formation}",
      journal = {\apj},
         year = 2007,
        month = apr,
       volume = {658},
       number = {2},
        pages = {1236-1251},
          doi = {10.1086/511513},
archivePrefix = {arXiv},
       eprint = {astro-ph/0611528},
 primaryClass = {astro-ph},
       adsurl = {https://ui.adsabs.harvard.edu/abs/2007ApJ...658.1236P}
}

@ARTICLE{LesurPapaloizou2010,
       author = {{Lesur}, G. and {Papaloizou}, J.~C.~B.},
        title = "{The subcritical baroclinic instability in local accretion disc models}",
      journal = {\aap},
         year = 2010,
        month = apr,
       volume = {513},
          eid = {A60},
        pages = {A60},
          doi = {10.1051/0004-6361/200913594},
archivePrefix = {arXiv},
       eprint = {0911.0663},
 primaryClass = {astro-ph.EP},
       adsurl = {https://ui.adsabs.harvard.edu/abs/2010A&A...513A..60L}
}

@ARTICLE{KlahrHubbard2014,
       author = {{Klahr}, Hubert and {Hubbard}, Alexander},
        title = "{Convective Overstability in Radially Stratified Accretion Disks under Thermal Relaxation}",
      journal = {\apj},
         year = 2014,
        month = jun,
       volume = {788},
       number = {1},
          eid = {21},
        pages = {21},
          doi = {10.1088/0004-637X/788/1/21},
archivePrefix = {arXiv},
       eprint = {1403.6721},
 primaryClass = {astro-ph.SR},
       adsurl = {https://ui.adsabs.harvard.edu/abs/2014ApJ...788...21K}
}

@ARTICLE{Lyra2014,
       author = {{Lyra}, Wladimir},
        title = "{Convective Overstability in Accretion Disks: Three-dimensional Linear Analysis and Nonlinear Saturation}",
      journal = {\apj},
         year = 2014,
        month = jul,
       volume = {789},
       number = {1},
          eid = {77},
        pages = {77},
          doi = {10.1088/0004-637X/789/1/77},
archivePrefix = {arXiv},
       eprint = {1405.3437},
 primaryClass = {astro-ph.EP},
       adsurl = {https://ui.adsabs.harvard.edu/abs/2014ApJ...789...77L}
}

@ARTICLE{Latter2016,
       author = {{Latter}, Henrik N.},
        title = "{On the convective overstability in protoplanetary discs}",
      journal = {\mnras},
         year = 2016,
        month = jan,
       volume = {455},
       number = {3},
        pages = {2608-2618},
          doi = {10.1093/mnras/stv2449},
archivePrefix = {arXiv},
       eprint = {1510.06247},
 primaryClass = {astro-ph.EP},
       adsurl = {https://ui.adsabs.harvard.edu/abs/2016MNRAS.455.2608L}
}

@ARTICLE{UrpinBrandenburg1998,
       author = {{Urpin}, V. and {Brandenburg}, A.},
        title = "{Magnetic and vertical shear instabilities in accretion discs}",
      journal = {\mnras},
         year = 1998,
        month = mar,
       volume = {294},
       number = {3},
        pages = {399-406},
          doi = {10.1046/j.1365-8711.1998.01118.x10.1111/j.1365-8711.1998.01118.x},
       adsurl = {https://ui.adsabs.harvard.edu/abs/1998MNRAS.294..399U}
}

@ARTICLE{Nelson+2013,
       author = {{Nelson}, Richard P. and {Gressel}, Oliver and {Umurhan}, Orkan M.},
        title = "{Linear and non-linear evolution of the vertical shear instability in accretion discs}",
      journal = {\mnras},
         year = 2013,
        month = nov,
       volume = {435},
       number = {3},
        pages = {2610-2632},
          doi = {10.1093/mnras/stt1475},
archivePrefix = {arXiv},
       eprint = {1209.2753},
 primaryClass = {astro-ph.EP},
       adsurl = {https://ui.adsabs.harvard.edu/abs/2013MNRAS.435.2610N}
}

@ARTICLE{Richard+2016,
       author = {{Richard}, Samuel and {Nelson}, Richard P. and {Umurhan}, Orkan M.},
        title = "{Vortex formation in protoplanetary discs induced by the vertical shear instability}",
      journal = {\mnras},
         year = 2016,
        month = mar,
       volume = {456},
       number = {4},
        pages = {3571-3584},
          doi = {10.1093/mnras/stv2898},
archivePrefix = {arXiv},
       eprint = {1601.01921},
 primaryClass = {astro-ph.EP},
       adsurl = {https://ui.adsabs.harvard.edu/abs/2016MNRAS.456.3571R}
}

@ARTICLE{Lesur+2025,
       author = {{Lesur}, Geoffroy and {Latter}, Henrik N. and {Ogilvie}, Gordon I.},
        title = "{High-resolution models of the vertical shear instability}",
      journal = {arXiv e-prints},
         year = 2025,
        month = aug,
          eid = {arXiv:2508.20839},
        pages = {arXiv:2508.20839},
          doi = {10.48550/arXiv.2508.20839},
archivePrefix = {arXiv},
       eprint = {2508.20839},
 primaryClass = {astro-ph.EP},
       adsurl = {https://ui.adsabs.harvard.edu/abs/2025arXiv250820839L}
}

@ARTICLE{Marcus+2013,
       author = {{Marcus}, Philip S. and {Pei}, Suyang and {Jiang}, Chung-Hsiang and {Hassanzadeh}, Pedram},
        title = "{Three-Dimensional Vortices Generated by Self-Replication in Stably Stratified Rotating Shear Flows}",
      journal = {\prl},
         year = 2013,
        month = aug,
       volume = {111},
       number = {8},
          eid = {084501},
        pages = {084501},
          doi = {10.1103/PhysRevLett.111.084501},
archivePrefix = {arXiv},
       eprint = {1303.4361},
 primaryClass = {astro-ph.EP},
       adsurl = {https://ui.adsabs.harvard.edu/abs/2013PhRvL.111h4501M}
}

@ARTICLE{SurvilleMayer2019,
       author = {{Surville}, Cl{\'e}ment and {Mayer}, Lucio},
        title = "{Dust-vortex Instability in the Regime of Well-coupled Grains}",
      journal = {\apj},
         year = 2019,
        month = oct,
       volume = {883},
       number = {2},
          eid = {176},
        pages = {176},
          doi = {10.3847/1538-4357/ab3e47},
archivePrefix = {arXiv},
       eprint = {1801.07509},
 primaryClass = {astro-ph.EP},
       adsurl = {https://ui.adsabs.harvard.edu/abs/2019ApJ...883..176S}
}

@ARTICLE{Meheut+2012b,
       author = {{Meheut}, H. and {Meliani}, Z. and {Varniere}, P. and {Benz}, W.},
        title = "{Dust-trapping Rossby vortices in protoplanetary disks}",
      journal = {\aap},
         year = 2012,
        month = sep,
       volume = {545},
          eid = {A134},
        pages = {A134},
          doi = {10.1051/0004-6361/201219794},
archivePrefix = {arXiv},
       eprint = {1208.4947},
 primaryClass = {astro-ph.EP},
       adsurl = {https://ui.adsabs.harvard.edu/abs/2012A&A...545A.134M}
}

@ARTICLE{Raettig+2015,
       author = {{Raettig}, Natalie and {Klahr}, Hubert and {Lyra}, Wladimir},
        title = "{Particle Trapping and Streaming Instability in Vortices in Protoplanetary Disks}",
      journal = {\apj},
         year = 2015,
        month = may,
       volume = {804},
       number = {1},
          eid = {35},
        pages = {35},
          doi = {10.1088/0004-637X/804/1/35},
archivePrefix = {arXiv},
       eprint = {1501.05364},
 primaryClass = {astro-ph.EP},
       adsurl = {https://ui.adsabs.harvard.edu/abs/2015ApJ...804...35R}
}

@Inbook{BenderOrszag1999,
      author="Bender, Carl M. and Orszag, Steven A.",
      title="Advanced Mathematical Methods for Scientists and Engineers",
      year="1999",
      publisher="Springer",
      address="New York, NY",
      pages="544--576",
      isbn="978-1-4757-3069-2",
      doi="10.1007/978-1-4757-3069-2_11",
      url="https://doi.org/10.1007/978-1-4757-3069-2_11"
}

@ARTICLE{LinYoudin2017,
       author = {{Lin}, Min-Kai and {Youdin}, Andrew N.},
        title = "{A Thermodynamic View of Dusty Protoplanetary Disks}",
      journal = {\apj},
         year = 2017,
        month = nov,
       volume = {849},
       number = {2},
          eid = {129},
        pages = {129},
          doi = {10.3847/1538-4357/aa92cd},
archivePrefix = {arXiv},
       eprint = {1708.02945},
 primaryClass = {astro-ph.EP},
       adsurl = {https://ui.adsabs.harvard.edu/abs/2017ApJ...849..129L}
}

@ARTICLE{Garaud+2004,
       author = {{Garaud}, P. and {Barri{\`e}re-Fouchet}, L. and {Lin}, D.~N.~C.},
        title = "{Individual and Average Behavior of Particles in a Protoplanetary Nebula}",
      journal = {\apj},
         year = 2004,
        month = mar,
       volume = {603},
       number = {1},
        pages = {292-306},
          doi = {10.1086/381385},
archivePrefix = {arXiv},
       eprint = {astro-ph/0307199},
 primaryClass = {astro-ph},
       adsurl = {https://ui.adsabs.harvard.edu/abs/2004ApJ...603..292G}
}

@ARTICLE{Nakagawa+1986,
       author = {{Nakagawa}, Y. and {Sekiya}, M. and {Hayashi}, C.},
        title = "{Settling and growth of dust particles in a laminar phase of a low-mass solar nebula}",
      journal = {\icarus},
         year = 1986,
        month = sep,
       volume = {67},
       number = {3},
        pages = {375-390},
          doi = {10.1016/0019-1035(86)90121-1},
       adsurl = {https://ui.adsabs.harvard.edu/abs/1986Icar...67..375N}
}

@ARTICLE{Shen+2006,
       author = {{Shen}, Yue and {Stone}, James M. and {Gardiner}, Thomas A.},
        title = "{Three-dimensional Compressible Hydrodynamic Simulations of Vortices in Disks}",
      journal = {\apj},
         year = 2006,
        month = dec,
       volume = {653},
       number = {1},
        pages = {513-524},
          doi = {10.1086/508980},
archivePrefix = {arXiv},
       eprint = {astro-ph/0609131},
 primaryClass = {astro-ph},
       adsurl = {https://ui.adsabs.harvard.edu/abs/2006ApJ...653..513S}
}

@ARTICLE{Dritschel1990,
       author = {{Dritschel}, David G.},
        title = "{The stability of elliptical vortices in an external straining flow}",
      journal = {Journal of Fluid Mechanics},
         year = 1990,
        month = jan,
       volume = {210},
        pages = {223-261},
          doi = {10.1017/S0022112090001276},
       adsurl = {https://ui.adsabs.harvard.edu/abs/1990JFM...210..223D}
}

\appendix
\section{The gas model}
\label{sec:gas_model}

Model A relies on an unconventional gas model that combines feature of compressible hydrodynamics (the gas density evolves over time), and of incompressible hydrodynamics (there is no equation of state, and pressure is a Lagrange multiplier enforcing a constraint on the divergence of the velocity field). The goal of this appendix is to derive this model.

\newcommand{\cs}{c_{\text{s}}} 

To do so, we start from the Navier-Stokes equations in the shearing box:
\begin{subequations}
    \label{eq:Compressible_Navier_Stokes_equations}
    \begin{align}
        & \partial_{t} \rhog + \bug \bcdot \bnabla \rhog = - \rhog \, \bnabla \bcdot \bug , \label{eq:Compressible_Navier_Stokes_equations_density} \\
        & \partial_{t} \bug + \bug \bcdot \bnabla \bug = - \frac{\bnabla P}{\rhog} - 2 \Omega \, \eZ \wedge \bug + 2 \Omega S X \eX , \label{eq:Compressible_Navier_Stokes_equations_momentum}
    \end{align}
\end{subequations}
and we assume an isothermal equation of state, ${ P = \cs^{2} \, \rhog }$.

\newcommand{\trhog}{\tilde{\rho}_{g}}
\newcommand{\tbug}{\tilde{\bu}_{g}}

\newcommand{\M}{\mathcal{M}} 

We then adimensionalise those equations using the orbital timescale ${ T = 1 / \Omega }$, an arbitrary lengthscale $L$ and an arbitrary density scale $\overline{\rho}$. This yields
\begin{subequations}
    \label{eq:Adimensional_compressible_Navier_Stokes_equations}
    \begin{align}
        & \partial_{\tt} \trhog + \tbug \bcdot \tbnabla \trhog = - \trhog \, \tbnabla \bcdot \tbug , \label{eq:Adimensional_compressible_Navier_Stokes_equations_density} \\
        & \partial_{\tt} \tbug + \tbug \bcdot \tbnabla \tbug = - \frac{\bnabla \th}{\trhog} - 2 \, \eZ \wedge \tbug + 2 \, (S / \Omega) \tilde{X} \, \eX , \!\! \label{eq:Adimensional_compressible_Navier_Stokes_equations_momentum} \\
        & \th = \trhog / \M^{2} , \label{eq:Adimensional_compressible_Navier_Stokes_equations_EoS}
    \end{align}
\end{subequations}
where ${ \M = \Omega L / \cs }$ is the Mach number of the flow. We shall assume that it is much smaller than one.

\renewcommand{\trhogo}{\tilde{\rho}_{g, 0}}
\newcommand{\trhogi}{\tilde{\rho}_{g, 1}}
\newcommand{\trhogii}{\tilde{\rho}_{g, 2}}

\newcommand{\tbugo}{\tilde{\bu}_{g, 0}}
\newcommand{\tbugi}{\tilde{\bu}_{g, 1}}
\newcommand{\tbugii}{\tilde{\bu}_{g, 2}}

Finally, we decompose each variable in powers of~$\M^{2}$:
\begin{align}
    \trhog &= \trhogo + \M^{2} \, \trhogi + \M^{4} \, \trhogii  + .., \nonumber \\
    \tbug &= \tbugo + \M^{2} \, \tbugi + \M^{4} \,\tbugii + .., \nonumber \\
    \th &= \M^{-2} \tho + \thi + \M^{2} \, \thii + ... \nonumber
\end{align}
We use a different scaling for $\th$ than $\trhog$ to compensate for the ${ \M^{2} }$ factor in the equation of state. At this stage it is important to note that there are other consistent scalings, leading to other models (Boussinesq, anelastic, \textit{etc.}). 

In fact, our scaling is only valid if $\tho$ and $\trhogo$ are uniform. Under this assumption, the leading-order equations are
\begin{subequations}
    \begin{align}
        & \tbnabla \bcdot \tbugo = - \partial_{\tt} \ln{ (\trhogo) }  , \phantom{\frac{1}{1}} \nonumber \\
        & \partial_{\tt} \tbugo + \tbugo \bcdot \tbnabla \tbugo = - \frac{\tbnabla \thi}{\trhogo} - 2 \, \eZ \wedge \tbugo + 2 \, (S / \Omega) \tilde{X} \, \eX , \nonumber \\
        & \trhogi = \M^{2} \, \thi . \phantom{\frac{1}{1}} \nonumber
    \end{align}
\end{subequations}

Note that we arranged the equations to make their triangular structure evident: the user provides the parameter ${ \partial_{\tt} \trhogo }$, which controls ${ \tbnabla \bcdot \tbugo }$, which controls the Lagrange multiplier $\thi$, which controls the evolution of $\tbugo$. If the user sets ${ \partial_{\tt} \ln{ (\trhogo) } = 0 }$, we recover the incompressible model.

Interestingly, $\thi$ also sets the value of $\trhogi$. This means that the second-order equation will have a similar structure, with pressure acting as a Lagrange multiplier. In fact, the same thing happens at all orders: the equation of state at order $n$ transforms the continuity equation at order ${ n + 1 }$ into a constraint, which can only be satisfied by using $\tilde{h}_{n+1}$ as a Lagrange multiplier. This introduces a closure problem.

Fortunately, it turns out that we can truncate the expansion at any order to obtain a closed system. For instance, if we truncate at leading order, forget the indices and bring back the dimensions, we get
\begin{subequations}
    \label{eq:Gas_model}
    \begin{align}
        & \bnabla \bcdot \bug = - \partial_{t} \ln{ (\rhog) }  , \phantom{\frac{1}{1}} \label{eq:Gas_model_continuity} \\
        & \partial_{t} \bug + \bug \bcdot \bnabla \bug = - \frac{\bnabla h}{\rhog} - 2 \Omega \, \eZ \wedge \bug + 2 \Omega S X \eX . \label{eq:Gas_model_momentum}
    \end{align}
\end{subequations}
This is the gas model that we use in the main text. It is valid for low-Mach-number flows with nearly uniform density. The gas density evolves over time, but there is no equation of state, and pressure is a Lagrange multiplier enforcing this density evolution.

Note that this truncation introduce an error of size ${ \mathcal{O} (\M^{2}) }$. On the other hand, the multiple-timescale methods of sections~\ref{sec:Model_0} and~\ref{sec:Model_AB} involve terms of order ${ \mathcal{O} (\St) }$. Therefore, our dusty vortex models are only rigorously valid if ${ \M^{2} \ll \St }$. This restrict our analysis to relatively small vortices with ${ L/H \ll \sqrt{\St} }$, where $L$ is the vortex's radial lengthscale and $H$ the disc's radial pressure scale height. To provide a point of comparison, global simulations routinely show vortices of size ${ L \sim H }$ (see, \textit{e.g;} \citealt{Shen+2006}).

\bsp	
\label{lastpage}
\end{document}